\documentclass[12pt,preprint]{aastex}

\def\ltsima{$\; \buildrel < \over \sim \;$}
\def\lsim{\lower.5ex\hbox{\ltsima}}
\def\gtsima{$\; \buildrel > \over \sim \;$}
\def\gsim{\lower.5ex\hbox{\gtsima}}

\shorttitle{Photometric Redshifts for GOODS-S Galaxies}
\shortauthors{Dahlen et al.}

\begin{document}


\title{A Detailed Study of Photometric Redshifts for GOODS-South Galaxies}


\author{Tomas Dahlen\altaffilmark{1}, 
Bahram Mobasher\altaffilmark{2},
Mark Dickinson\altaffilmark{3},
Henry C. Ferguson\altaffilmark{1},
Mauro Giavalisco\altaffilmark{4},
Norman A. Grogin\altaffilmark{1},
Yicheng Guo\altaffilmark{4},
Anton Koekemoer\altaffilmark{1},
Kyoung-Soo Lee\altaffilmark{5},
Seong-Kook Lee\altaffilmark{6},
Mario Nonino\altaffilmark{7},
Adam G. Riess\altaffilmark{6},
and Sara Salimbeni\altaffilmark{4}
}
\email{dahlen@stsci.edu}
  
\altaffiltext{1}{Space Telescope Science Institute, 3700 San Martin Drive, Baltimore, MD 21218, USA}
\altaffiltext{2}{Department of Physics and Astronomy, University of California, Riverside, CA 92521, USA}
\altaffiltext{3}{NOAO, 950 N. Cherry Avenue, Tucson, AZ 85719, USA}
\altaffiltext{4}{Department of Astronomy, University of Massachusetts, Amherst, MA 01003, USA}
\altaffiltext{5}{Yale Center for Astronomy and Astrophysics, Departments of Physics and Astronomy, Yale University, New Haven, CT 06520, USA}
\altaffiltext{6}{Department of Physics and Astronomy, Johns Hopkins University, 3400 North Charles Street, Baltimore, MD 21218, USA}
\altaffiltext{7}{INAF - Osservatorio Astronomico di Trieste, Via Tiepolo 11, I-34131 Trieste, Italy}

\begin{abstract}
We use the deepest and the most comprehensive photometric data currently available for GOODS-South (GOODS-S) galaxies to measure their photometric redshifts. The photometry includes VLT/VIMOS ($U$~band), $HST$/ACS (F435W, F606W, F775W, and F850LP bands), VLT/ISAAC ($J$, $H$, and $K_s$~bands), and four $Spitzer$/IRAC channels (3.6, 4.5, 5.8, and 8.0$\mu$m). The catalog is selected in the $z$~band (F850LP) and photometry in each band is carried out using the recently completed TFIT algorithm, which performs point-spread function (PSF) matched photometry uniformly across different instruments and filters, despite large variations in PSFs and pixel scales. Photometric redshifts are derived using the GOODZ code, which is based on the template fitting method using priors. The code also implements ``training'' of the template spectral energy distribution (SED) set, using available spectroscopic redshifts in order to minimize systematic differences between the templates and the SEDs of the observed galaxies. Our final catalog covers an area of 153 arcmin$^2$~and includes photometric redshifts for a total of 32,505 objects. The scatter between our estimated photometric and spectroscopic redshifts is $\sigma$=0.040 with 3.7\% outliers to the full $z$-band depth of our catalog, decreasing to  $\sigma$=0.039 and 2.1\% outliers at a magnitude limit $m_z<$24.5. This is consistent with the best results previously published for GOODS-S galaxies, however, the present catalog is the deepest yet available and provides photometric redshifts for significantly more objects to deeper flux limits and higher redshifts than earlier works. Furthermore, we show that the photometric redshifts estimated here for galaxies selected as dropouts are consistent with those expected based on the Lyman break technique.
\end{abstract}


\section{INTRODUCTION}

A major focus in observational astronomy in recent years has been the study of changes in the intrinsic properties of galaxies (i.e., absolute luminosity, stellar mass, star formation rate, radial size etc.) with look-back time. This requires redshifts to a large number of galaxies detected to deep flux levels. Such galaxy surveys have recently become possible with the advent of sensitive panoramic detectors with large field of view, capable of detecting the faintest objects. This hints toward the need for techniques to measure accurate redshifts to a large number of faint galaxies. The most accurate and convincing way to do this is by performing spectroscopic observations. However, constructing galaxy surveys based on spectroscopic redshifts is severely limited by the large number of galaxies for which redshifts are needed, by their flux limit, and partly by the absence of emission lines in optical spectra (i.e., the redshift desert, $1.4\lsim z \lsim 2.5$). As a result, spectroscopic surveys are biased against fainter galaxies, suffer from the spectroscopic selection effects (i.e., whether the spectroscopic sample is magnitude, color or surface brightness limited) and, over a range of redshifts, the method fails when the observed spectral lines from the targets are contaminated by sky emission lines (i.e., redshift desert). To avoid the above problems, we exploit photometric redshift techniques, allowing us to measure redshifts to large number of galaxies to very faint flux levels, with a relatively modest investment in observing time. 

In recent years, photometric redshifts have played a pivotal role in studying different aspects of galaxy evolution. This includes studies of the evolution, with look-back time, of the luminosity function (Dahlen et al. 2005), mass function (Ilbert et al. 2010), star formation rate (Gabasch et al. 2004; Dahlen et al. 2007) the large-scale structure (Scoville et al. 2007) in the universe, and in identifying high-redshift candidates (Steidel et al. 1996). Moreover, they are used in selecting sources for follow-up spectroscopic observations, in confirming spectroscopic redshifts that are only based on a single line (Lilly et al. 2009), and in identifying high-redshift galaxy clusters (Adami et al. 2010).

The photometric redshift techniques can be divided into two broad categories, based on template fitting (e.g., HyperZ (Bolzonella et al. 2000); BPZ (Ben\'{i}tez 2000); ImpZ (Babbedge et al. 2004); ZEBRA (Feldmann et al. 2006); LePhare (S. Arnouts \& O. Ilbert 2010, in preparation); EAZY (Brammer et al. 2008); Low Resolution Template (LRT) Libraries (Assef et al. 2008); GALEV (Kotulla et al. (2009)) and empirical codes (ANNz (Collister \& Lahav 2004); Multilayer Perceptron Artificial Neural Network (Vanzella et al. 2004); Gaussian Process Regression (Way \& Srivastava 2006); ArborZ (Gerdes et al. 2010); ``Empirical-$\chi^2$'' (Wolf 2009); ``Random Forests'' (Carliles et al. 2010)). In the former method, a set of observed or synthetic spectral energy distributions (SEDs) is used to predict the expected magnitudes and colors of the objects of interest at different redshifts. These are then compared with the observed SEDs of galaxies and their respective redshifts, corresponding to the best-fitting solution, are estimated via $\chi^2$ minimization. The latter technique uses a training set of galaxies with known redshifts to derive a relation between the observed fluxes and spectroscopic redshifts and then applies this to the objects without spectroscopic redshifts to deriving their photometric redshifts. No template SEDs are required in this method, but a large spectroscopic sample is needed for ``training'' purposes. Recently, ``hybrid'' codes have also been developed which are able to train template SEDs (e.g., Ilbert et al. 2006). For comparisons between different publicly available codes, see Abdalla et al. (2008) and Hildebrandt et al. (2008, 2010).

The accuracy of photometric redshifts depends, to a large extent, on the spectral coverage and resolution (number of available photometric bands) of the SEDs of galaxies in question. The main source of uncertainty is the photometric errors, especially at faint magnitudes where the signal-to-noise ratio (S/N) is low (Dahlen et al. 2008). Also, in multi-waveband surveys, magnitudes are measured with different telescopes and instruments. Differences in the point-spread functions (PSFs), corresponding to different instrument and seeing conditions lead to fluxes measured over different physical areas of a galaxy, compromising the accuracy of their SEDs. Furthermore, in constructing multi-waveband catalogs, using data with very different PSFs, we face the blending problem, where images taken with instruments with broader PSFs or bad seeing conditions merge and hence, making photometry (and in some cases even source identification) difficult. 

In this paper we measure photometric redshifts for galaxies in the Great Observatories Origins Deep Survey southern field (GOODS-S). There are a number of major improvements here compared to previous measurements. Firstly, the photometric redshifts are estimated for a catalog selected in Advance Camera for Surveys (ACS) $z$~band. This has the highest spatial resolution ($\sim$ 0.03 arcsec) and is significantly deeper than previous catalogs. Secondly, the photometry is performed using the new template fitting (TFIT; Laidler et al. 2007) technique where high-resolution images are matched to images with significantly lower resolution. This reduces photometric uncertainties due to blending and provides consistent photometry in all the passbands regardless of the PSF size. A code using the same concept for deriving photometry (ConvPhot, De Santis et al. 2007) has previously been used by Grazian et al. (2006) to derive photometric redshifts for GOODS-S, see Section 5.1. Thirdly, we measure photometric redshifts for ACS detected sources, exploiting data extended to mid-infrared wavebands. Including mid-infrared channels in photometric redshift fitting has recently become more feasible by including imaging conducted by $Spitzer$/IRAC (e.g., Grazian et al. 2006; Wuyts et al. 2008).    

The paper is organized as follows: In Section 2, we describe the GOODS-S photometric and spectroscopic data used in this study. Our photometric redshift code (hereafter GOODZ-- Dahlen et al. 2005; Mobasher et al. 2007) is presented in Section 3. In Section 4 we estimate photometric redshifts for galaxies in the GOODS-S field, for which spectroscopic redshifts are available and explore their dependence on the choice of filters, templates and magnitude limit. In Section 5 we compare our results with those from literature, and in Section 6 we apply our photometric redshift technique to galaxies selected through the dropout techniques. The GOODS-S photometric redshift catalog is described in Section 7. We summarize our results in Section 8. Throughout, we assume a cosmology with $\Omega_{\rm M}$=0.3, $\Omega_{\Lambda}$=0.7, and $h$=0.7. Magnitudes are given in the AB system.     
   
\section{DATA AND THE PHOTOMETRIC CATALOG}

\subsection{Photometric Data}

In the present study, we concentrate on the GOODS-S field. The data used in this paper were collected as a part of the GOODS campaign\footnote{http://www.stsci.edu/science/goods/} and are briefly explained below.

{\it U-band Data:} the $U$-band data were obtained using the VIMOS (Le F\`{e}vre et al. 2003) instrument on Very Large Telescope (VLT). The observations cover a total area of 630 arcmin$^2$~and reach a S/N=5 depth for point sources $m_{\rm AB}$=28.0 mag for the area common to the GOODS-S $HST$/ACS observations. The FWHM of the $U$-band data is 0.8 arcsec (Nonino et al. 2009). These data are significantly deeper than the CTIO and ESO/WFI $U$-band data used in previous studies of the GOODS-S field (Giavalisco et al. 2004a; Mobasher et al. 2004; Dahlen et al. 2007). Moreover, it is matched in sensitivity to the data in other passbands. Note that the VIMOS $U$-band filter, as well as the CTIO $U$~band, is non-standard with a redder transmission compared to a standard Johnson $U$~band.

{\it Optical Data}: this consists of imaging data in four $HST$/ACS passbands: F435W, F606W, F775W, and F850LP, hereafter referred to as the $B$, $V$, $i$ and $z$~band, respectively. This covers the entire GOODS-S area of $10\times 16$ arcmin$^2$. The imaging observations were carried out during five GOODS epochs (Giavalisco et al. 2004a). These images were then combined with further ACS exposures from the $HST$~Supernovae search project (Riess et al. 2007), adding five more epochs, making these among the deepest $HST$~data currently available. The FWHM of the PSF is $\sim$0.11 arcsec for the ACS data. The final S/N=5 depths for point sources (using a circular aperture radius of 0.25 arcsec) of the GOODS-S version 2.0 $HST$~ACS images are $m_{\rm AB}$=28.7, $m_{\rm AB}$=28.8, $m_{\rm AB}$=28.3, and $m_{\rm AB}$=28.1 in $B$, $V$, $i$, and $z$~band, respectively. For extended sources (using an aperture radius 0.25 arcsec), the S/N=5 limits are $m_{\rm AB}$=28.0, $m_{\rm AB}$=28.0, $m_{\rm AB}$=27.5, and $m_{\rm AB}$=27.3 in the four ACS bands. The ACS images have been drizzled to a resolution of 0.03 arcsec pixel$^{-1}$~from the original plate scale of 0.05 arcsec pixel$^{-1}$.

{\it Near-Infrared (NIR) Data:} the $J$-, $H$- and $K_s$-band data were obtained using the Infrared Spectroscopic And Array Camera (ISAAC) on the VLT (Retzlaff et al. 2010). These data have a PSF ranging from 0.34 to 0.65 arcsec and cover a major fraction of the GOODS-S area ($>$80\% in all bands). The $5\sigma$~point source limiting magnitudes in $J$, $H$, and $K_s$~bands are $m_{\rm AB}$=25.0, $m_{\rm AB}$=24.5 and $m_{\rm AB}$=24.4 (within 75\% of the area covered), respectively (Retzlaff et al. 2010). We use the final data release version 2.0. 

{\it Mid-Infrared (MIR) Data:} the mid-infrared data were obtained as a part of the GOODS $Spitzer$~campaign, using the InfraRed and Array Camera (IRAC) instrument. The data were taken in all the four IRAC channels centered on wavelengths 3.6, 4.5, 5.8, and 8.0 $\mu$m. Hereafter, we refer to these as ch1, ch2, ch3, and ch4. The sampling here is 0.6 arcsec, much worse than that for the $HST$~ACS data. The FWHM of the PSF is $\sim$1.7 arcsec in ch1, ch2, and ch3 and $\sim$1.9 arcsec in ch4. The final IRAC data have magnitude limits (S/N=5) for point sources corresponding to $m_{\rm AB}$=26.1 (ch1), $m_{\rm AB}$=25.5 (ch2), $m_{\rm AB}$=23.5 (ch3) and $m_{\rm AB}$=23.4 (ch4).

\begin{figure}
\epsscale{0.9}
\plotone{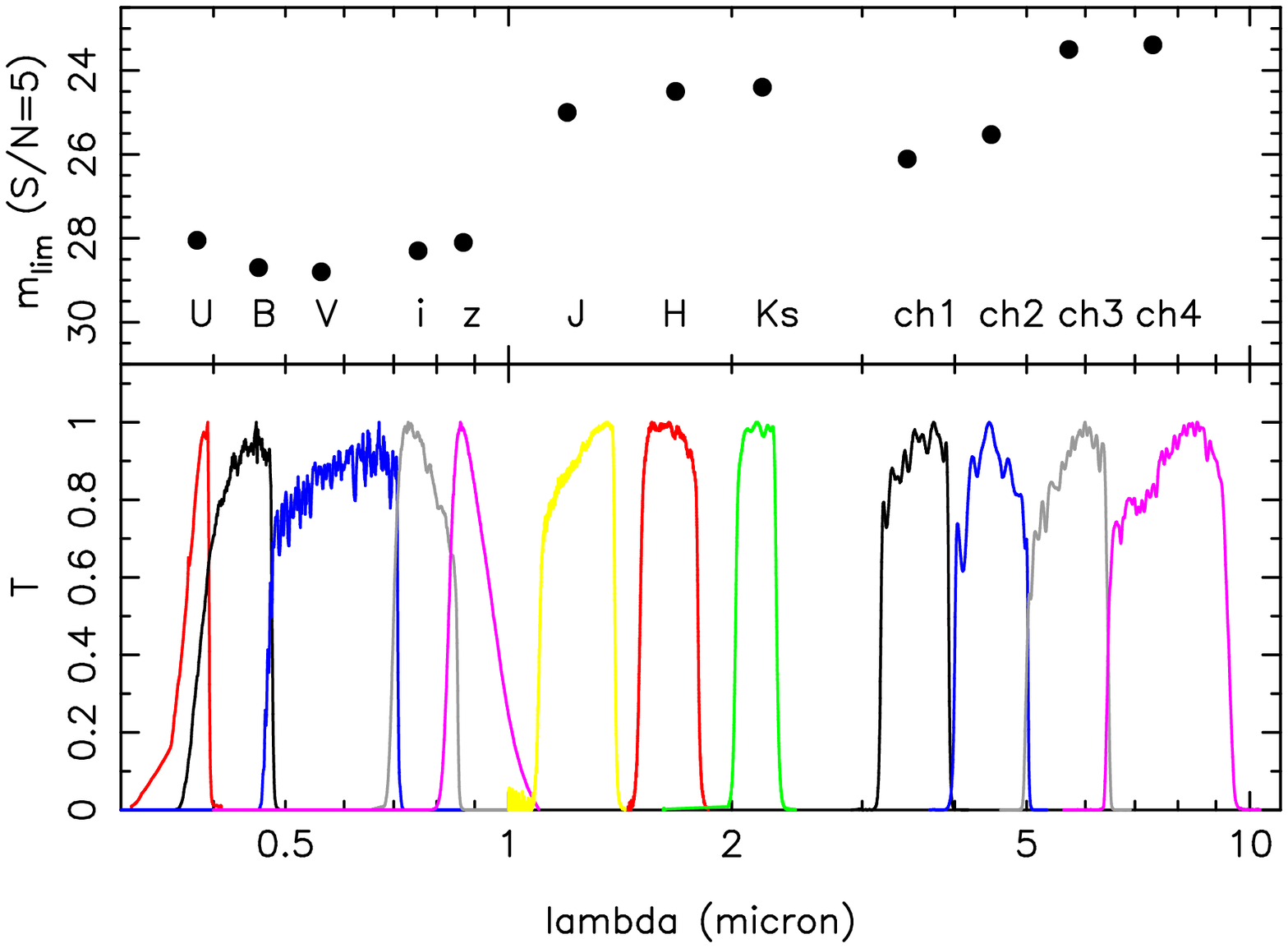} 
\caption{Bottom panel shows filter transmissions functions with maximum transmission normalized to unity in each filter, while the top panel shows the S/N=5 point source limiting magnitudes for each filter.
\label{fig1}}
\end{figure}

In Figure \ref{fig1}, we show the S/N=5 limiting magnitudes for point sources together with the bandpasses for the 12 filters used in this investigation. Maximum transmission is normalized to unity in each filter.

\subsection {Multi-waveband Photometric Catalog}
Given the various data set we have, the main challenge in performing the photometry is to generate a multi-waveband catalog with consistent photometry, using data with vastly different pixel scales (from 0.03 arcsec (ACS) to 0.6 arcsec (IRAC)) and PSFs ($\sim$0.1 arcsec to $\sim$2 arcsec). The large PSF size of the IRAC images leads to serious blending, causing problem with source identification (i.e., sources detected and resolved with ACS may be blended at IRAC wavelengths). Moreover, this causes photometric uncertainties due to contamination by neighboring sources. 

Our photometric catalog is based on source extraction performed on the ACS $z$-band data using the SExtractor software (Bertin \& Arnouts 1996). This is the deepest of the included images with high spatial resolution (0.03 arcsec). We then use the TFIT technique (Laidler et al. 2007) to perform matched photometry between the data at different passbands with widely different PSFs. The TFIT code uses the positions and two-dimensional surface brightness profiles of sources measured at high resolution, in this case the $HST$/ACS $z$-band, as priors to measuring their fluxes at lower resolution in other passbands. The TFIT code recovers accurate fluxes even when high source density causes significant blending at lower resolutions, particularly in the case of GOODS $Spitzer$/IRAC and ground-based NIR images. The final ACS $z$-band-selected photometric catalog covers an area of 153 arcmin$^2$~and includes a total of 32,508 objects to a (S/N)$_z=$3 limit (81\% of the objects have (S/N)$_z>$5). For a detailed discussion of the multi-waveband TFIT catalog, we refer the reader to N. Grogin et al. (2010, in preparation). 

To compare results based on the TFIT and conventional photometry, we also produce SExtractor (Bertin \& Arnouts 1996) catalogs. First, we put the ACS $B$, $V$, $i$, and $z$~images (0.03 arcsec pixel$^{-1}$) on the same pixel scale as the ISAAC $J$, $H$, and $K_s$~data (0.15 arcsec pixel$^{-1}$) by block-summing the former using 5$\times$5 pixels and preserving the fluxes. Images are thereafter smoothed to a common PSF. We then create a $z$-band-selected catalog with magnitudes from the four ACS and three ISAAC bands, derived using SExtractor in ``dual image mode''. Finally, we match this catalog by coordinates to independent SExtractor catalogs of VIMOS $U$~band and IRAC photometry, using a matching radius of 0.5 arcsec. For the reminder of this paper we call this the ``SExtractor catalog''. Because of the way the SExtractor catalog is generated, there is not a one-to-one correspondence between the galaxies in that catalog and the TFIT catalog (i.e., individually resolved galaxies in the TFIT catalog may be blended in the SExtractor catalog).

\subsection {Spectroscopic Data}
The spectroscopic redshift catalog for GOODS-S was compiled from a large number of sources (Cristiani et al. (2000); Croom et al. (2001); Dickinson et al. (2004); LeF\`{e}vre et al. (2004); Stanway et al. (2004); Strolger et al. (2004); Szokoly et al. (2004); van der Wel et al. (2004); Doherty et al. (2005); Mignoli et al. (2005); Roche et al. (2006); Ravikumar et al. (2007); Popesso et al. (2009); Vanzella et al. (2008); D. Stern et al. in preparation (2010)). For each redshift reference, object coordinates were matched to the GOODS ACS version 2.0 $z$-band catalog and, where necessary, a net offset was applied to bring the coordinate systems into average agreement. The nearest cataloged ACS source to the adjusted literature position was then adopted as the counterpart, up to a maximum matching radius of 0.5 arcsec. No other matching criteria were used in the counterpart matching. However, if two or more spectroscopic redshift measurements were available from the literature for a given ACS galaxy, several considerations were used to decide which value would be adopted, or whether the object would be discarded from the photometric redshift comparisons altogether. First, most of the spectroscopic surveys attach a set of quality flags to their redshift measurements, indicating their nominal level of reliability. We remapped these different quality flag systems to a single three-level scale. Quality flag 1 indicates a secure redshift, flag 2 is a probable redshift with some chance of error due to misidentification of spectral features, and flag 3 implies an insecure redshift. For redshifts from the GOODS ESO spectroscopic surveys (Vanzella et al. 2008; Popesso et al. 2009) and those from D. Stern et al. (2010, in preparation), this was a straight remapping of their A/B/C quality scale. The remapping for other surveys was somewhat subjective, but in any case, as we will describe below, objects with contentious redshift measurements were excluded from the photometric redshift. The quality flags only indicate the reported quality of the redshift measurements, and not the reliability of the matching to the counterparts in the photometric catalog. After remapping, the quality flags were compared for cases with multiple redshift observations, and the measurement with the highest quality was adopted. If more than one survey yielded a match with equal quality flag values, these were chosen according to an order of precedence based on previous experience concerning the overall reliability of the redshift data.  In the end, 2875 objects from our photometric catalog were assigned spectroscopic redshifts with any quality, of which 172 are stars.

\begin{figure}
\epsscale{0.7}
\plotone{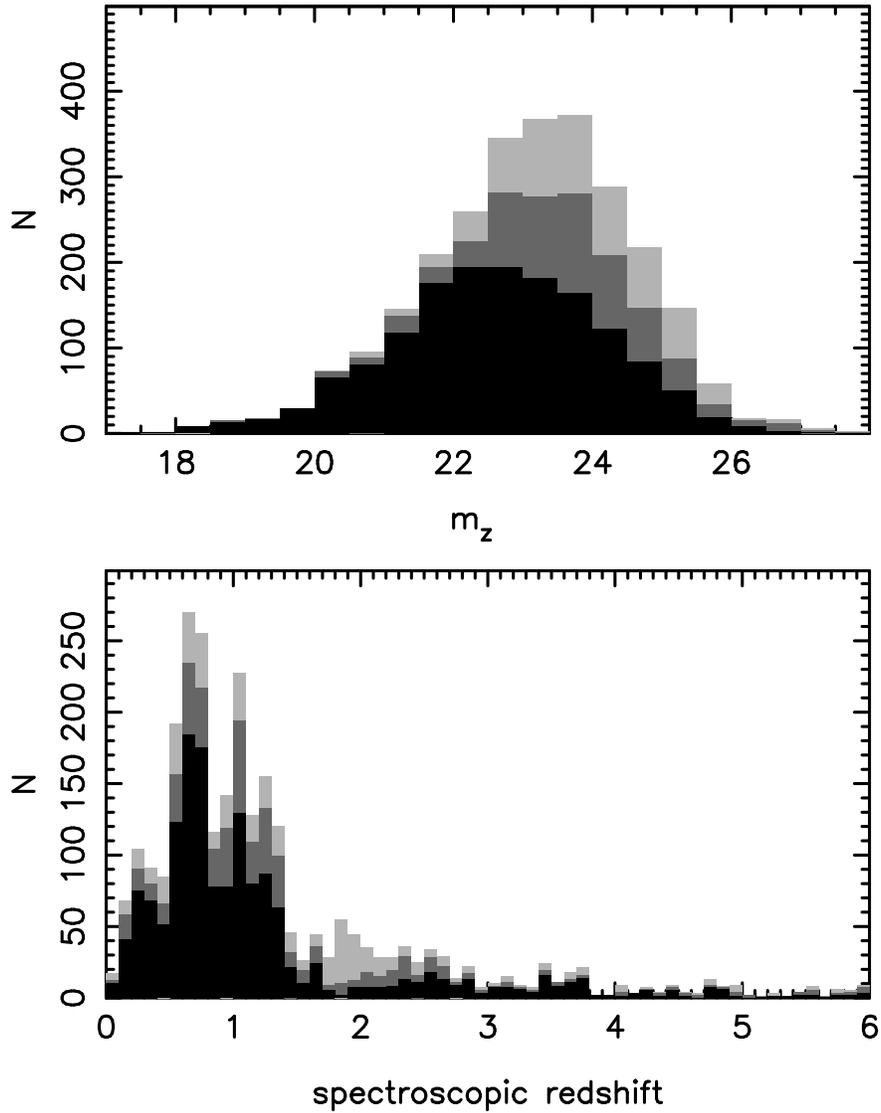}
\caption{Histogram showing the magnitude distribution in the $z$~band for available spectroscopic redshifts (top panel) and the redshift distribution of the spectroscopic redshifts (bottom panel) for GOODS-S. Black color shows objects with quality flag=1, while dark and light gray colors represent additional objects with quality flag=2 and 3, respectively.
\label{fig2}}
\end{figure}

In Figure \ref{fig2}, we plot the redshift and magnitude distributions of the GOODS-S spectroscopic sample. In the plot, objects with quality flag=1 are shown with black color, while additional objects with quality flag=2 and 3 are shown with dark gray and light gray colors, respectively. The numbers of non-stellar objects in the three subsamples are 1534, 608, and 561, respectively. 

Out of the 2875 objects matching to the spectroscopic redshift catalogs, 657 have multiple spectra available.  Out of these, there are 188 cases where spectroscopic measurements disagree by more than $| \Delta z | > 0.02$ (considering redshifts of all quality classes). Later in this study, when we compare photometric and spectroscopic redshifts, we adopt a very conservative strategy, including only objects with quality flag~$=$~1 and excluding any objects with multiple spectroscopic redshifts that are discrepant by more than $| \Delta z | > 0.02$. This leaves a sample of 1403 objects with spectroscopic redshifts used for such comparisons.

\section{PHOTOMETRIC REDSHIFTS}
We derive photometric redshifts using the GOODZ code, which is primarily a template fitting method  (e.g., Gwyn 1995; Mobasher at al. 1996) and is a developed version of the code earlier applied to GOODS-S data (see Dahlen et al. 2005; 2007 for more details). This method compares the observed fluxes for a given galaxy with those from template SEDs, shifted in redshift space. At each redshift a $\chi^2$~value is assigned by minimizing
\begin{equation}
\chi^2(z)=\sum^n_{i=1}([F^i_{\rm obs}-\alpha F^i_{\rm template}]/\sigma^i)^2,
\end{equation} 
where $n$ is the number of passbands available and $F^i_{\rm obs}$~and $F^i_{\rm template}$~are, respectively, the observed and template fluxes in any given band, $i$. Here, $F^i_{\rm template}$~includes information on the template SEDs for different galaxy types, internal and intergalactic absorption, as well as the response curves for the filters. Finally, $\alpha$~is a normalization constant and $\sigma^i$~is the flux error in $F^i_{\rm obs}$. We require $n\ge$3, i.e., a detection or an upper limit in at least three filters, in order to calculate photometric redshifts.

We use template SEDs from Coleman et al. (1980) covering types E, Sbc, Scd, Im, and two starburst SEDs from Kinney et al. (1996; templates SB2 and SB3). The Coleman et al. templates are extended to UV and IR wavelengths as described in Bolzonella et al. (2000). Besides these six discrete templates, we use four interpolations between each pair of templates from early to late types. This provides a set of 26 discrete templates. In addition, to each template we also apply eight discrete extinction corrections in the range $E(B-V)=-0.1$~to $E(B-V)=0.3$ with steps $\Delta E(B-V)=0.05$. For early type to spiral galaxies, we use the Cardelli et al. (1989) extinction law, while for later type spirals and starbursts we use the Calzetti et al. (2000) extinction law. Therefore, at each redshift we minimize $\chi^2$~using a set of 234 discrete templates. In redshift space, each template SED is shifted in the range $0<z<7$~in steps $\Delta z=0.01$. We therefore use a total of 164,034 discrete template SEDs. Intergalactic absorption is treated using the recipe in Madau (1995). The code also includes algorithms for flagging stars and other point source objects as further described in Sections 3.1 and 4.4.

\begin{figure}
\epsscale{0.9}
\plotone{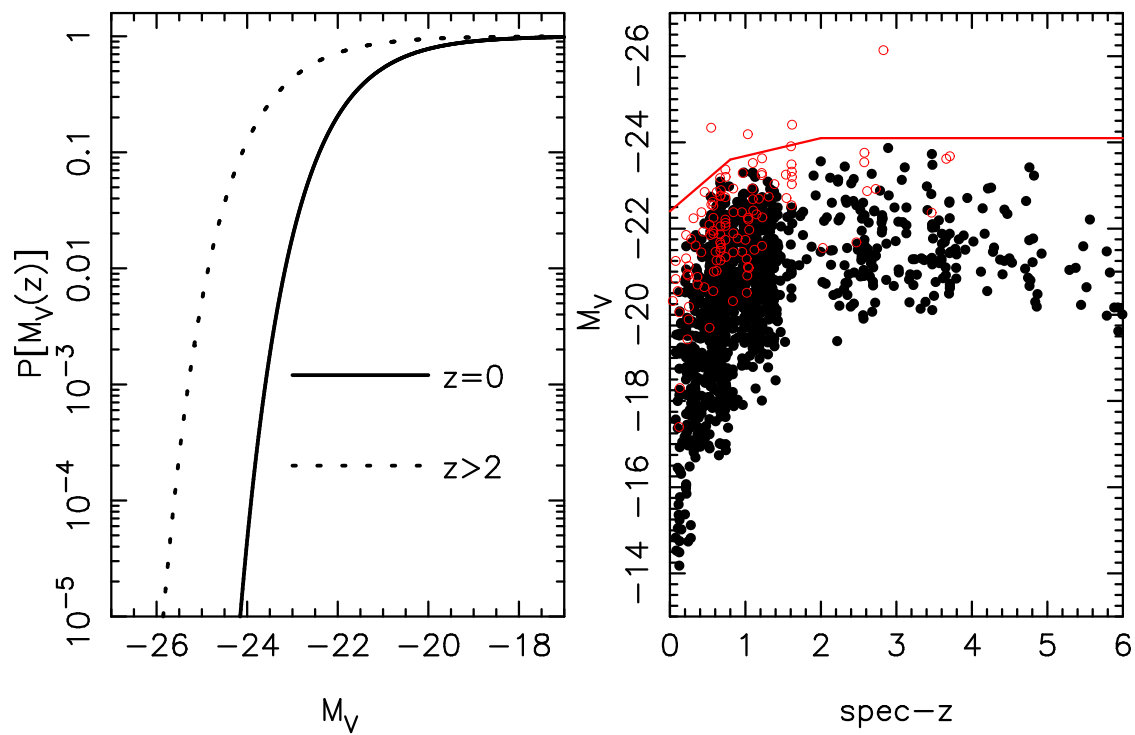}
\caption{Left panel shows the probability function of the adopted prior on the absolute magnitude distribution. The prior evolves with redshift from $z=0$~to $z=2$, thereafter it stays constant. Right panel shows the absolute magnitudes for the spectroscopic sample. All normal galaxies (black dots) are below the line that shows where the prior probability function $P[M_V(z)]=0.1$. A few X-ray sources are brighter than this limit (open circles).
\label{fig3}}
\end{figure}

We include a luminosity function prior in the photometric redshift fitting (e.g., Kodama et al. 1999; Ben\'{i}tez 2000). The idea is to convolve the redshift probability distribution derived from the $\chi^2$~fitting with a probability distribution based on priors. For the prior distribution, we calculate the absolute $V$-band magnitude the galaxy would have at each tested redshift and compare this with that most probably expected from the input luminosity function. If the absolute magnitude corresponding to a particular redshift is improbable, i.e., significantly brighter than $M^*$, then this redshift is disfavored, i.e., a low value is assigned to the probability distribution. Given the shape of a Schechter luminosity function, this works as an exponential cutoff at bright magnitudes. Taking into account the brightening of $M^*$~with redshift, we use a cutoff that gets 1.7 mag brighter from $z=0$~to $z=2$ and thereafter stays constant. At faint limits, $M_V\gsim M^*$, we assume that all absolute magnitudes are equally probable. We denote the probability of a galaxy having an absolute magnitude of $M_V$~at redshift $z$~by $P$[$M_V(z)$]. In the left panel of Figure \ref{fig3}, we show the probability distribution as a function of absolute magnitude at $z=0$~and $z>2$. The right panel shows the absolute magnitudes for the spectroscopic sample. The location where the probability is $P$[$M_V(z)$]=0.1 is shown with the line. All normal galaxies, shown as black dots, are found below this line (i.e., at $P$[$M_V(z)$]$>0.1$). Only a few X-ray sources (circles) have magnitudes brighter than the line. This approach of having an absolute magnitude prior is similar to that used by Rowan-Robinson et al. (2008), except that they use a single cutoff magnitude at each redshift, while we use a probability distribution at each redshift. We have chosen to evolve $M^*$~by an amount that both matches Figure \ref{fig3}~and is consistent with observations suggesting a brightening of $M^*$~with redshift in optical bands, including the $V$~band used here (e.g., Ilbert et al. 2005). With the prior probability function given by $P$[$M_V(z)$], our final estimated probability distribution is
\begin{equation}
P(z)\propto {\rm exp}(-\chi^2(z))*P[M_V(z)]
\end{equation}
We thereafter define the photometric redshift
\begin{equation}
z_{\rm phot}= \frac{\int zP(z)dz}{\int P(z)dz}.
\end{equation}
In the case that there are two or more discrete peaks in the redshift probability distribution, the integration is performed over the main (highest) peak. Besides $z_{\rm phot}$~we also calculate the value at the peak of the probability distribution, $z_{\rm peak}$. In the results section, we compare the photometric redshifts based on these two quantities. Furthermore, we also calculate the 68\% and 95\% confidence intervals for the photometric redshifts using the full $P(z)$~distribution.

There are a number of sources that can introduce systematic errors in the template fitting method if used ``blindly''. In particular, the template SEDs may not adequately represent the true shapes and range of the galaxy SEDs, and the spectral shapes may evolve with redshift. Also, possible zero-point errors in the photometry introduce systematics that may bias results. Finally, if the estimated photometric errors are too small (or large), the weighting by the error in Equation (1) may bias the fit toward (away from) some of the filters, which may also cause systematic effects. However, with a large number of spectroscopic redshifts available, it is possible to correct for some of these biases. For example, Ilbert et al. (2006) used spectroscopic redshifts to correct both for zero-point offsets and adjust the template SEDs used in the fitting procedure. Here, we use a similar method, described in more detail below.

\subsection{Point Sources}
\begin{figure}
\epsscale{0.7}
\plotone{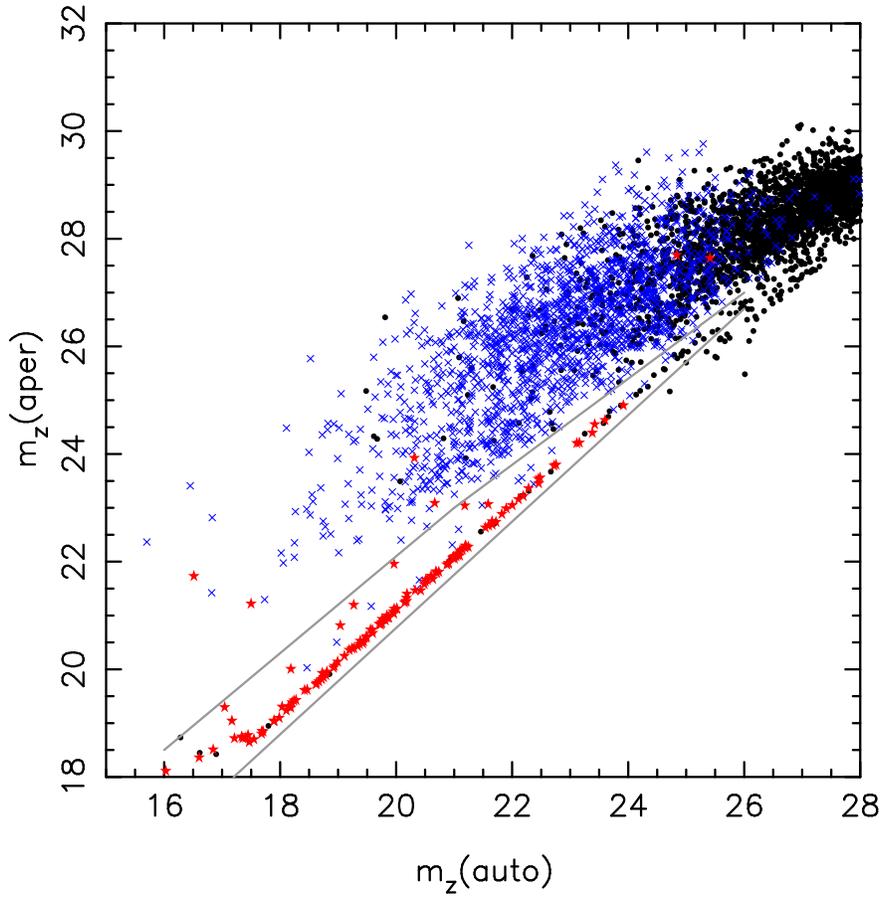}
\caption{SExtractor aperture magnitude vs. SExtractor MAG$_{\rm AUTO}$~(corresponding to the total) in the $z$~band. The stellar sequence is clearly visible. The asterisks are spectroscopically confirmed stars. Crosses are non-stellar objects. Dots are sources with no available spectroscopic redshifts. 
Straight lines represent our selection criterion for identifying the point sources. 
\label{fig4}}
\end{figure}

The GOODZ code uses special algorithms for flagging point source objects such as stars and QSOs based on morphology and color selection. To morphologically select point sources, we plot, in Figure \ref{fig4}, the $z$-band SExtractor aperture magnitude (using an aperture corresponding to the seeing FWHM) versus the SExtractor MAG$_{\rm AUTO}$~which closely corresponds to total magnitude. This is equivalent to a surface brightness versus magnitude plot. In this figure, objects that are spectroscopically identified as stars are shown with asterisks, while non-stellar objects with spectroscopic redshifts are shown with crosses. In addition, a random selection of objects without spectroscopic redshifts is shown with dots. The sequence of point sources is clearly visible in the figure. We use the lines shown in the figure to locate the locus of the point sources. In addition, we also require $m_z<$26 in order to flag an object as a point source, since at fainter magnitudes the increased photometric errors as well as the smaller sizes of objects make the selection less accurate, with an increased possibility of objects scattering in and out of the point source selection locus. In Figure \ref{fig4}, there are six stellar objects ($z_{\rm spec}$=0.0; asterisk) outside the region identified by the point source selection criteria. Visual inspection shows that four of these have close companions and the remaining two are saturated, both circumstances affecting the location on the plot.
\subsubsection{Stellar Colors}
\begin{figure}
\epsscale{0.7}
\plotone{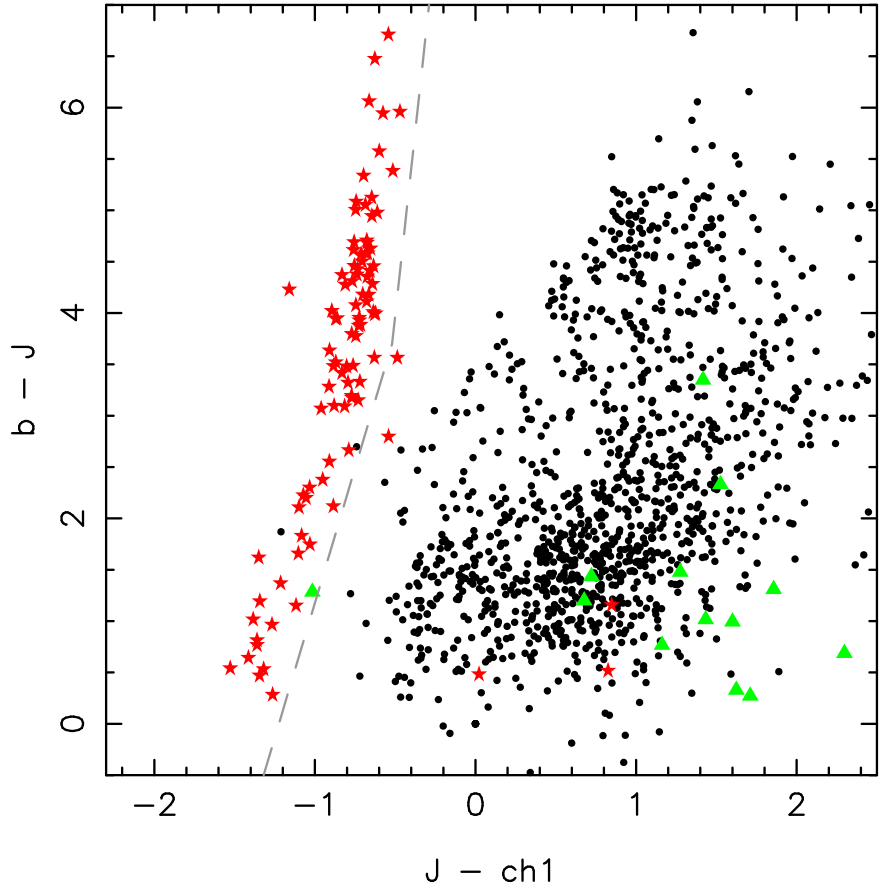}
\caption{Color--color ($b-J$) vs. ($J-$ch1) diagram for objects with high quality spectroscopic redshifts. The asterisks show spectroscopically confirmed stars while triangles show point source objects (i.e., AGNs) excluding stars. Remaining dots are non-point source galaxies.
\label{fig5}}
\end{figure}

To separate stars, we use a ($b-J$) versus ($J-$ch1) color--color selection criterion (Mancini et al. 2009). In Figure \ref{fig5}, we show the observed color--color diagram for 1511 sources with high quality spectroscopic redshifts. A subsample of 128 spectroscopically identified stars is shown with asterisks. It is evident from this plot that these objects form a tight sequence separated from the colors of normal galaxies. In the GOODZ code, we use a color--color selection indicated by the dashed line in Figure \ref{fig5} to flag objects that are consistent with being stars. While we still calculate the photometric redshifts for all objects in our catalog, this flag can be used to separate objects that most probably are stars.

There are three objects in Figure \ref{fig5} that meet the color criterion for being stars (located left of selection line) but have spectroscopic redshifts different from $z_{\rm spec}$=0. The top two of these objects (reddest $b-J$~colors) have spectroscopic redshifts $z_{\rm spec}$=3.791\footnote{R.A.=53.1388588, decl.=--27.8353806} and $z_{\rm spec}$=3.484\footnote{R.A.=53.1048164, decl.=--27.8146114} from Vanzella et al. (2008). A visual inspection shows that neither of the objects have a point source morphology but that both objects have a close neighbor (within 1 arcsec). It is likely that the companion objects have affected the colors, so that they fall within the stellar sequence (the close companion could also have affected spectroscopic redshift determination). The third object falling within the star selection criterion has a spectroscopic redshift $z_{\rm spec}$=0.086\footnote{R.A.=53.1580276, decl.=--27.7691936} from Szokoly et al. (2004). Inspecting the image shows a double system with one point source close to ($\sim$0.5 arcsec) a second, fainter, point source. We therefore believe that the object is a star and that the non-zero redshift may be due to a close companion. 

Besides these three objects, there are also five objects with $z_{\rm spec}$=0, but with colors in the color--color diagram that are outside the stellar sequence in Figure \ref{fig5}. Three of these objects have non-points source morphologies, inconsistent with being stars\footnote{R.A.=53.0598457, decl.=--27.7849779 - nearby star; R.A.=53.1592641, decl.=--27.9359062; R.A.=53.1683313, decl.=--27.8767088}. For one object there is, however, a nearby star ($\sim$2 arcsec), suggesting a possible error in the catalog matching. For the remaining two non-point source objects, the morphologies and colors suggest that the spectroscopic redshifts are probably wrong. Finally, two objects with $z=0.0$~and point source morphologies fall outside the selection criterion\footnote{R.A.=53.0620162, decl.=--27.7725833; R.A.=53.0703781, decl=--27.8420718}. These objects are the ones that are closest to the dashed line in Figure \ref{fig5} (distance in magnitude $\Delta m$=0.08 and 0.17) and should therefore be consistent with the star selection used considering the photometric errors. We could have relaxed the selection to include these objects by shifting the line to redder ($J-$ch1) colors. However, that leads to an increased risk of flagging a larger number of ``non-stellar'' objects in our final catalog. In the results section, we give the total number of objects that are flagged as stars.
  
After safely excluding the three non-point source objects as not being stars, we note that our color--color selection criterion correctly selects 123 of the remaining 125 objects. 

\subsubsection{QSOs}
While we can flag stars using color--color selections, this is not the case for the QSOs. We could, in principle, also include a set of templates covering different types of QSOs and active galactic nuclei (AGNs) in our spectral template library. However, this will lead to increased risk of degeneracy when fitting normal galaxies without AGN contribution, resulting in an increased fraction of outliers. We therefore do not include additional templates to account for these SED types. However, we have conducted extensive tests where we investigate how the inclusion of extra templates affects the results as discussed and quantified in Section 4.4.

\subsection{Optimization of Photometric Redshifts using a Spectroscopic Training Sample} 
By using the sample of spectroscopic redshifts described above, we estimate and correct for systematic effects in the photometric redshift determination. This includes correcting for: (1) zero-point offsets in the photometry; (2) systematic biases in the photometric errors; (3) template SED deviations from observed SEDs; and (4) dependence of the template SEDs on redshift. In the first step, we use the GOODZ code and find the best-fit template for each galaxy after fixing the redshift to its known spectroscopic redshift. From this fit, we calculate the flux of the best-fitting SED in each observed filter. This flux is thereafter compared with the observed flux. Deviations between the template and observed flux can be attributed to flux errors in the observed flux, zero-point errors, calibration errors, insufficient knowledge of the filter response functions and deviations in the shape of the template SEDs from the ``true'' SEDs. For a single comparison, it is of course not possible to determine which of these effects are in play. However, with sufficiently large sample of spectroscopic redshifts, an iterative approach can be applied to estimate the contributions from different parts, as discussed below.

\subsubsection{Step 1 -- Zero-point Offsets}
To determine if there are any biases in the observed fluxes, for each filter, we plot in Figure \ref{fig6} the histogram of the magnitude differences between the observed photometry and that predicted by the best-fitting SED template. Any significant offset in the median of this distribution is likely to be due to a zero-point problem, deviations between the template SEDs and observed fluxes, or possible contributions from other effects such as incorrect aperture corrections. For this comparison, we use objects with spectroscopic redshifts of quality 1 and 2. Even though there may be a few incorrect redshifts in the latter category, by using the median as a measurement of the shift, the effect of such objects would not be significant.

In Figure \ref{fig6}, we show the offsets for all filters before any corrections as gray histograms (note the different scaling on the $x$-axis for the IRAC bands). For clarity we show the offsets in magnitude space, even though the actual fitting is done in flux space. Positive offsets in the figure indicate that the measured fluxes are brighter than the template fluxes. There are definitely biases in the observed fluxes. The offsets for the $U$~band and ACS bands are relatively small (at most $\sim$0.05 mag in the $B$~band), increasing somewhat for the ISAAC bands and IRAC ch1 and ch2 ($\sim$0.1 mag). Significant offsets are noted in IRAC ch3 and ch4 where fluxes are $\sim$0.3--0.5 mag too bright compared to the templates. There is a systematic offset between the instruments, particularly ISAAC and IRAC, indicating that there may be a problem when directly comparing fluxes from different instruments where e.g., pixel scales and PSFs differ. However, the relatively large offsets in IRAC ch3 and ch4 may partly be due to dust emission in the MIR regime not accounted for in the template SEDs. We discuss this further in Section 3.2.3

As a first step in optimizing the photometric redshifts, we correct the flux in each filter by the median offsets in the distributions shown by the gray histograms in Figure \ref{fig6} and in Table \ref{table1}. After this first correction, we recalculate the offsets and find that the median shifts have mostly disappeared after applying a zero-point correction to the flux. For the optical, the NIR bands, and IRAC ch1 and ch2 the median shift is typically 0.01, while for the IRAC bands ch3 and ch4 the shifts are $\lsim$0.02--0.04.

\subsubsection{Step 2 -- Systematic Biases in the Photometric Errors}
To account for systematic errors not included in the statistical flux errors assigned, we add, in quadrature, 0.05 mag in the optical bands, 0.1 mag in the ISAAC bands, and 0.20 mag in the IRAC channels to the existing statistical errors (corresponding to 5\%, 10\%, and 20\% additional flux errors, respectively). These ``smoothing'' errors are further discussed in the Appendix.
 
\subsubsection{Step 3 -- Optimizing the Template SEDs}
Next, we use the offsets between the observed fluxes and those predicted from template SED fluxes, after applying the offsets calculated above, to investigate rest-wavelength dependent systematics. For each photometric point, we plot in Figure \ref{fig7} the offset at the rest-frame wavelength: 
\begin{equation}
\lambda_{\rm rest} = \lambda_0/(1+z_{\rm spec})
\end{equation}
where $\lambda_0$~is the effective wavelength of the filter for which the offset is calculated and $z_{\rm spec}$~is the spectroscopic redshift. The offsets are shown separately for the six different template SEDs used. In each panel, we also plot the median offset, which at each point is calculated using 100 measurements. The median is plotted to $\sim 5\mu$m ($\sim 4\mu$m for SB galaxies), at longer wavelengths there are too few data points to derive the median. Both from the measurements themselves and the median, it is clear that there are systematic offsets between the template SEDs and the actual measured photometry. This effect is most evident at long wavelengths $\gsim3\mu$m. We note that the templates we have adopted are extended to MIR wavelengths, as described in Bolzonella et al. (2000), but that they do not include emission from dust in the interstellar medium (ISM). As an initial precaution, we have not included IRAC ch4 in the photometric fitting at $z<0.5$ since here this filter may be affected by the strong polycyclic aromatic hydrocarbon (PAH) feature at rest frame $\sim7.7\mu$m. This means that only rest-frame wavelengths $\lsim$6.5$\mu$m are used in the fitting. However, the results in Figure \ref{fig7} show that, in particular for the later type galaxies, the observed fluxes at long wavelengths are brighter than the template fluxes. Since late-type galaxies are more likely to have a significant contribution from dust emission, the trends seen in Figure \ref{fig7} are consistent with an underestimate of the true galaxy flux in the MIR caused by the lack of dust emission in the templates. The particular features affecting the photometry at the wavelengths $>3\mu$m include the PAH features at 3.3$\mu$m and 6.2$\mu$m. For the starburst templates, the underestimate of the galaxy flux starts at somewhat shorter wavelengths, $\sim2\mu$m, indicating that these templates may be under representing the flux already at NIR wavelengths. Note that a similar trend is seen by Brammer et al. (2008), who find a large offset between observed and template fluxes at MIR wavelengths and also a significant offset at NIR wavelengths.

To account for the differences between the observed and template fluxes, we construct a new set of template SEDs where we correct the original SEDs using the measured median offsets versus wavelength shown in Figure \ref{fig7}. At long wavelengths, $\sim 4-5\mu$m to 6.5$\mu$m, we assume that the corrections are constant with a value given by the last measured median data point for each template. Note that this procedure makes the choice of input set of template SEDs less important. If we had started with a different template set, the corrections would have made the result similar to what found here. For a detailed description on how to reconstruct galaxy template SEDs using observed photometry of a spectroscopic sample, see Budav\'{a}ri et al. (2000).

\subsubsection{Step 4 -- Redshift Dependence}
Having corrected the fluxes for median offsets between observed and template fluxes (Section 3.2.1) and optimized the template SEDs (Section 3.2.3), we now consider redshift dependences. In Figure \ref{fig8}, we plot the offset between observed and template fluxes as a function of redshift. Since we include offsets from all available filters for each object, we expect the median offset to be close to zero. The plot verifies this where the median is shown to be near zero at all redshifts. More interestingly, we also study the evolution of the rms between template and observed fluxes. In a scenario were the true SEDs of galaxies evolve with redshift and are no longer represented by the template set used in the fitting, we expect the rms, shown as the upper curve in Figure \ref{fig8}, to increase with redshift. There is a weak trend of an increasing rms with redshift by ~0.07 mag over the redshift range investigated. However, at the same time we note that the statistical photometric errors increase from a mean 0.06 mag at $z<1$~to a mean 0.12 mag at $2<z<3$, which is consistent with the increase in rms between the observed photometry and the template SEDs. We therefore conclude that there is no significant indication of a redshift evolution in the SEDs of the observed galaxies that is not represented in the templates used. There is possibly an indication of an increased scatter at redshifts $z>2.8$. However, at these redshifts the Lyman break becomes the most important feature for determining photometric redshifts. Since this feature is independent of galaxy type, deviations between the true galaxy SEDs and the template SEDs used in the fitting become less important. Note that at redshifts $z>3$, the statistics are too poor to extend this investigation. Based on the results from this study, we do not consider any redshift-dependent corrections to the library of template SEDs.
\subsubsection{Step 5 -- Iterating Zero-point Offsets}
Next, we recalculate the zero-point offsets after having applied the corrections from steps 1 and 3 above. The additional offsets are typically $\sim$0.01 mag for the $U$, ACS and ISAAC bands and $\sim$0.02 for the IRAC bands.

\subsubsection{Step 6 -- Final Iteration}
As a final step, we run the GOODZ code in ``training mode''. In this mode we rerun the photometric redshift code after adding additional offsets corresponding to $\pm$0.01 mag to the fluxes in each band. After a first run through all filters, we keep the offsets that result in a decrease in the scatter between spectroscopic and photometric redshifts compared to the case with no additional offsets. We thereafter make a second iteration adding $\pm$0.01 mag to the already derived offsets. The iterations are repeated until no improvement in the scatter is achieved. The additional shifts resulting from this training do not exceed 0.05 mag in any of the bands, except IRAC ch4 where the offset is 0.08 mag. The black curves in Figure \ref{fig6} show the resulting offset between the observed and the template magnitudes in each filter after applying magnitude shifts and using the corrected template SEDs. The figure shows that the offsets have been corrected and that the widths of the distributions have narrowed (higher peaks). 

Note that the shifts we have applied (Table \ref{table1}) were derived to optimize the photometric redshifts, which does not necessarily indicate that the photometry in the GOODS-S catalogs are affected by the same offsets. This is because there are many possible sources contributing to the offsets besides pure zero-point errors, including template mismatches, aperture corrections, errors in filter functions, etc. Therefore, we recommend that these offsets be not applied to the GOODS-S photometry for other purposes than when deriving photometric redshifts.
\section{RESULTS}
After applying the derived offset corrections to fluxes and using the updated set of template SEDs, we use the GOODZ code to derive photometric redshifts for the full sample of GOODS-S galaxies. To estimate the accuracy of the photometric redshifts, we compare them with their spectroscopic counterparts, using the subsample with a spectroscopic quality flag=1. To quantify the accuracy of photometric redshifts, we define
\begin{equation}
\sigma_z={\rm rms}[\Delta z/(1 + z_{\rm spec})].
\end{equation}
where $\Delta z=z_{\rm spec}-z_{\rm phot}$.

``Catastrophic'' redshifts outliers are defined as objects with $|\Delta z|/(1+z_{\rm spec})>$0.15 and we denote the scatter after excluding outliers by $\sigma_{zc}$. Furthermore, we define the bias as 
${\rm bias}_z={\rm mean}[(\Delta z/(1 + z_{\rm spec})]$.

As a second indicator of the photometric redshift quality, we use the normalized median absolute deviation of $\Delta z$~as given by
\begin{equation}
\sigma_{\rm NMAD}=1.48\times {\rm median}(\frac{|\Delta z-{\rm median}(\Delta z)|}{1+z_{\rm spec}}).
\end{equation}
This representation is less affected by the outliers and has recently been used in a number of surveys (e.g., Brammer et al. 2008; Ilbert et al. 2009; Pell\'{o} et al. 2009; Luo et al. 2010).

To calculate the deviation between the spectroscopic and photometric redshifts, we use 1118 spectroscopic redshifts with data quality=1, not including stars and AGN. For a default setup, using all 12 available bands and a magnitude limit $m_z<$24.5, we derive a scatter $\sigma_{zc}$=0.039 after excluding 2.1\% outliers. We apply this magnitude limit to exclude the faintest objects for which photometric errors are largest. However, this only excludes 13\% of the sample. For completeness, we note that scatter marginally increases, $\sigma_{zc}$=0.040 when including all magnitudes (1\,280 objects), however, the number of outliers increases more significantly to 3.7\%. For the full sample, without excluding outliers, the scatter is $\sigma_z$=0.062 and $\sigma_z$=0.135 for the two selections, respectively. Finally, for the normalized median absolute deviation we derive $\sigma_{\rm NMAD}$=0.034 ($m_z<$24.5) and  $\sigma_{\rm NMAD}$=0.035 (all magnitudes). This shows that the fraction of outliers increases somewhat when including the faintest objects while the scatter after excluding outliers is not highly dependent on magnitude.

The resulting scatter is plotted in the left panel of Figure \ref{fig9}. Black dots and crosses show objects brighter and fainter than $m_z$=24.5, respectively. The right panel shows the distribution of residuals between spectroscopic and photometric redshifts ($z_{\rm spec} - z_{\rm phot}$), with a best-fit Gaussian distribution with $\sigma$=0.056 overplotted. For the redshift normalized residuals ([$z_{\rm spec} - z_{\rm phot}$]/[1+$z_{\rm spec}$]), we get $\sigma$=0.038. Furthermore, we find bias$_z=-0.005$~and bias$_z=-0.006$~(after excluding outliers) for the $m_z<$24.5 sample and the full sample, respectively. Results on the scatter and fraction of outliers for different selections are also given in Table \ref{table2}. Figure \ref{fig9} indicates that there may be a bias in the redshift interval $2<z<3$ where the photometric redshifts are systematically lower than the spectroscopic redshifts. For this redshift range alone, we find bias$_z-0.014$~and bias$_z-0.032$~for the $m_z<$24.5 sample ($N=44$) and the full sample ($N=82$), respectively. Even though the statistics are relatively small, these results indicate that there may be increased uncertainties when calculating photometric redshifts in this redshift range.

We finally note that before making any corrections to the magnitudes or templates, the scatter was $\sigma_{zc}$=0.044 and $\sigma_{zc}$=0.045 for the two selections (with 2.9\% and 4.0\% outliers, respectively). This shows that the scatter and the fraction of outliers decreased as a result of the process described above.
\subsection{Sensitivity of Photometric Redshifts to Dust Emission} 

As noted above, the template set we use does not initially include the contribution due to dust emission in the ISM. Using the method described in Section 3.2.3, we adjust the template SEDs to account for differences between the observed flux and the template flux. For late-type galaxies, these corrections are consistent with dust emission at $\gsim 3\mu$m. When deriving the photometric redshifts, we make use of the corrected template SEDs, allowing us to include the IRAC bands in the photometric redshift fitting. However, to further examine the effect of dust emission, we also calculate the photometric redshifts where we include the mid-infrared bands in the SED fitting process only when they probe rest-frame wavelengths $ < 3\mu$m. This explicitly excludes the contribution due to dust when fitting the SEDs. Here, after recalculating the filter offsets as presented in Table \ref{table1}, we find that they mostly disappear for ch3 and ch4, suggesting that these offsets are indeed caused by MIR dust emission. However, this does not improve the accuracy of the photometric redshifts as measured in the last sections, with the scatter between the photometric and spectroscopic redshifts remaining the same. 

Finally, we examine the effects of dust on the photometric redshifts by adding dust emission to our templates using the SEDs from Ilbert et al (2009), which have incorporated dust emission at MIR wavelengths. We do, however, note that the inclusion of dust emission features to the template SEDs, and accurate modeling of it for different spectral types of galaxies is difficult and uncertain. Moreover, the PAH features are also sensitive to metallicity of their host galaxy (Calzetti et al. 2007), making reliable interpretation of these difficult. Furthermore, we find large differences between different dust models, with their relative contribution to the SEDs for different spectral types not known. Nevertheless, when using these templates, the observed offset is reduced but, again, the errors in the estimated photometric redshifts remain the same as before. 

Given that none of the above procedures improve the accuracy of our estimated photometric redshifts, we choose not to include the dust into our original template SEDs. Instead, we rely on the method described in Section 3.2.3 to correct our templates to account for dust emission.

The reddening applied to the template SEDs in the photometric redshift fitting lies in the range $-0.1<E(B-V)<0.3$. To investigate if the width of the allowed range affects results, we have also run the GOODZ code after expanding the range by a factor 3 to $-0.3<E(B-V)<0.9$. We find no significant differences when expanding the range and therefore conclude that the range used is sufficient for our purposes.

\subsection{Dependence on Redshift, Magnitude, and Color}
When determining the scatter between the spectroscopic and the photometric redshifts, the result will depend on the characteristics of the spectroscopic sample, i.e., redshift, magnitude, and color distributions. To examine how the results depend on these properties, we divide our sample in both redshift, magnitude, and color space. In Figure \ref{fig10}, we show the scatter as black dots (scaling on left $y$-axis) and the outlier fraction as histogram (scaling on right $y$-axis) as a function of magnitude in the interval $20<m_z<25.5$, using magnitude bins with size $\Delta m$=0.5. The figure shows that the scatter is fairly independent of magnitude while the fraction of outliers shows an increase at fainter magnitudes. The increase in outlier fraction is expected since fainter objects in general have higher spectroscopic redshifts and have larger photometric errors, which increases the risk for misidentifications of spectral features, leading to catastrophic redshifts.  

Next, we investigate the scatter versus redshift behavior by dividing our spectroscopic sample into redshift bins using a magnitude limit $m_z<24.5$. Since there are significantly fewer objects at high redshifts, we let the bin-size increase at higher redshifts. The black dots in Figure \ref{fig11} show that there is a trend of a slight increase in the scatter to $z\sim$2 (scaling on left-hand $y$-axis). However, in the higher redshift bins, the scatter is reduced to the same level as at low redshifts. This is due to shifting of the Lyman break feature into our observed passbands, making it easily detectable at these high redshifts. Horizontal error bars in the figure represent the bin-sizes. The fraction of outliers, shown by the histogram and scaling on right-hand side $y$-axis, increases with redshift. 

Finally, we investigate the scatter as a function of galaxy color. To quantify the galaxy color, we use the rest-frame ($B-V$) color of the best-fitting galaxy template after fixing the redshift of the template SEDs at the spectroscopic redshift. There is a strong color magnitude trend in the spectroscopic sample where the faintest galaxies are predominantly the bluest. Black dots in Figure \ref{fig12} show the scatter as a function of rest-frame ($B-V$) color for galaxies with $m_z<24.5$. The figure indicates that the photometric redshift accuracy do not depend strongly on galaxy color. Even though one may expect a higher scatter for later types or bluer colors do to the less pronounced 4000 \AA~break in these galaxies, this is not evident from the figure. One should, however, note that the spectroscopic sample includes relatively more blue galaxies at higher redshifts. Since the accuracy of the photometric redshifts increases at high redshifts once it is possible to detect the Lyman break, this helps the determination of the photometric redshifts for the population of blue galaxies.

To investigate if there is a redshift-color bias between the spectroscopic redshift sample and the photometric redshift sample, we calculate the fraction of galaxies of different spectral types that have redshifts below and above $z=2$, i.e., the redshift where the bluest filter ($U$~band) starts to probe the redshifted Lyman-break. For the $z<2$~subsample, we find that the spectroscopic sample ($N_{\rm spec}$=1140) consists of 13.1\% early-types, 38.0\% late-type spirals, and 48.9\% starbursts. For the photometric redshift sample ($N_{\rm phot}$=4798), the fractions are 12.3\%, 35.9\%, and 51.8\%, respectively. The composition of the samples agrees well at low redshift. For the high-redshift subsample with $z>2$, we find 0\% early-types, 1.4\% late-type spirals, and 98.6\% starbursts for the spectroscopic sample ($N_{\rm spec}$=70). For the photometric redshift sample ($N_{\rm phot}$=253), we find fractions 0\%, 1.2\%, and 98.8\%, respectively. The agreement is good also in this subsample, even though the statistics are smaller. We conclude that there is a trend of an increased fraction of starburst galaxies at higher redshift, but that there is no apparent bias in the fraction of spectral types between the spectroscopic redshift sample and the photometric redshift catalog. 
 In this example, we assign a starburst type to galaxies with ($B-V$)$<$0.34 and an early type for galaxies with ($B-V$)$>$0.66, while intermediate color galaxies are called late-types. Note, however, that highly obscured star-forming galaxies may be assigned an early galaxy type by this one color classification. But since there are relatively few early types galaxies, especially at high redshift, we do not expect that this has a significant effect on the relative abundances. These divisions are indicated in Figure \ref{fig12}. The only pronounced color trend in the figure is the higher fraction of outliers at the bluest colors. Inspecting this population, we find that 89\% of the outliers have z$_{\rm spec}>2$, indicating that a misidentification between the Lyman break and the 4000 \AA~break contributes to the higher outlier fraction at the bluest colors.

\subsection{Dependence on Filter Availability}
It is well known that the wavelength baseline covered by the filters used is important for the accuracy of the derived photometric redshifts. In particular, the $U$~band is important for a secure detection of the Balmer break at low redshifts and for detecting the Lyman break features at $z>$~2. Having NIR bands is crucial for detecting the Balmer break at $z\gsim$1.2 and therefore, for the accuracy of the photometric redshifts out to $z\sim 2.5-3$~until the Lyman break moves into the optical band. Some recent surveys (e.g., Grazian et al. 2006; Wuyts et al. 2008) have started to include the IRAC bands when calculating photometric redshifts, but further evaluation of the effect MIR bands on the photometric redshift accuracy is important. Below, we test all these scenarios in order to quantify their importance for photometric redshift measurements.

\subsubsection{$U$~Band}
The overall scatter in photometric redshifts increases from $\sigma_{zc}$=0.039 to $\sigma_{zc}$=0.044 when excluding the $U$~band in the template fitting, which is a fairly moderate increase. The effect is larger at lower redshifts ($z<0.3$), since here the $U$~band is most important for locating the Balmer break. For 94 objects with $z<0.3$, we derive a scatter $\sigma_{zc}$=0.043 when using all filters. This increases to $\sigma_{zc}$=0.064 when excluding the $U$~band. Also, the fraction of outliers increases dramatically from 2.1\% to 13.8\%. This stresses the importance of deep $U$~band data for local and low-redshift surveys. 

The $U$~band should also be particularly important in the redshift range where the filter probes rest-frame wavelengths short of the Lyman break, before the $B$~band moves into this break. For the VIMOS $U$~band filter used here, the redshift range where the $U$~band is the only filter probing the Lyman break is $2.0\lsim z\lsim 2.3$. For the 12 objects with spectroscopic redshifts and $m_z<24.5$ in this range, we find $\sigma_{zc}$=0.049 and 0\% outliers when including the $U$~band. After excluding the $U$~band, scatter becomes $\sigma_{zc}$=0.075 with 17\% outliers. Even though the statistical sample is small, these results indicate that the $U$~band is important at redshifts $z\sim2$.

\subsubsection{Infrared Passbands}
Having a long wavelength baseline is important for accurate measurement of photometric redshifts. Many recent investigations have shown that including infrared data is crucial for reducing the photometric redshift scatter. This is particularly important at $z\gsim1.2$~where the optical bands move to rest-frame wavelengths short of the Balmer break. At even higher redshifts ($z>2$), the $U$~and optical bands start to probe the Lyman break that to some extent again gives more secure redshifts. However, the short baseline when not including IR bands increases the risk for catastrophic failures (e.g., misidentification between Balmer and Lyman breaks). Also, this decreases the possibility to break the degeneracy between galaxy type and intrinsic reddening. Excluding the IR data, both ISAAC and IRAC, increases the scatter from $\sigma_{zc}$=0.039 to $\sigma_{zc}$=0.047 while, at the same time, the number of outliers more than doubles from 2.1\% to 5.5\%. Furthermore, if we restrict our sample to redshifts $z>1.2$~(223 objects) where the infrared bands become most important since the optical bands no longer straddle the 4000 \AA~break, the scatter increases from  $\sigma_{zc}$=0.044 to $\sigma_{zc}$=0.051 while the number of outliers increases from 4.5\% to 9.0\% when excluding the infrared bands. 

Excluding the ISAAC data while retaining the IRAC data has a relatively smaller effect on the scatter since this still keeps the long baseline. This results in only a very marginal increase in the scatter from $\sigma_{zc}$=0.039 to $\sigma_{zc}$=0.041. Likewise, excluding the IRAC bands while retaining the ISAAC data has a marginal effect, also resulting in an overall scatter of $\sigma_{zc}$=0.041. When we exclude the ISAAC or IRAC data, the fraction of outliers increases from 2.1\% to 2.5\% and 2.6\%, respectively. The relatively low impact on the photometric redshifts when including the IRAC data could be attributed to the fact that, at the redshift range of our interest, the ISAAC bands already provide a tight constraint on the SED long-wards of the Balmer break. 

\subsubsection{$z_{\rm phot}$~Versus $z_{\rm peak}$} 
Besides deriving the effective photometric redshift, $z_{\rm phot}$, by integrating the probability distribution according to Equation (3), we also list in our catalog the redshift at the peak of the probability distribution, $z_{\rm peak}$. We have already shown that for $z_{\rm phot}$, the scatter is $\sigma_{zc}$=0.040 with 3.7\% and $\sigma_{zc}$=0.039 with 2.1\% outliers for the full sample and for $m_z<$24.5, respectively. For the peak photometric redshift, $z_{\rm peak}$, the corresponding numbers are $\sigma_{zc}$=0.043 (both selections) with 3.9\% and 2.2\% outliers. This shows that the effective photometric redshift results in somewhat smaller scatter and lower outlier fraction and is therefore preferred to be used compared to the peak photometric redshift.

\subsection{Redshifts for Point Sources}
\subsubsection{Stars}
We flag stars using a color--color selection as described in Section 3.1.1. In total, we find 845 objects with colors consistent with being stars in our full catalog. These objects are flagged in our photometric redshift catalog, although photometric redshifts are assigned to these objects using galaxy SED templates. Of these flagged objects, there are 115 sources with spectroscopic redshifts consistent with being stars. There are also a total of 464 objects with point source morphologies according to our selection criteria described in the next section. About 57\% of these objects (264) have colors consistent with being stars.

\subsubsection{X-ray Sources} 
The template set we use does not include specific templates for X-ray sources like QSOs and AGNs. Furthermore, when calculating the scatter, we exclude known X-ray sources. To investigate if the inclusion of such sources affects results, we have matched our catalog with the Chandra X-ray catalog covering GOODS-S (Alexander et al. 2003). Of our 1209 objects with spectroscopic flag=1 and $m_z<$24.5, we find that 91 non-stellar objects have X-ray detection (for the full magnitude range the number of matched objects is 136). Deriving the scatter when including the X-ray sources does not increase the scatter ($\sigma_{zc}$=0.039), however, the fraction of outliers increases from 2.1\% to 2.9\% when including X-ray sources. The scatter of the X-ray subsample alone is $\sigma_{zc}$=0.048 after excluding 13.2\% outliers. This shows that the template set used here, also fits the SEDs for X-ray detected galaxies reasonably well. However, the fraction of outliers is higher for the X-ray sample, with 12 out of the 91 X-ray sources being outliers. This means that even though only $\sim$8\% of the objects with spectroscopic redshifts have X-ray flux, $\sim$34\% of the outliers belong to this category. 

The CDF-S X-ray catalog by Alexander et al. (2003) contains a total of 326 sources, of which $\sim$270 are inside the area covered by the GOODS-S ACS-$z$-selected catalog. Matching these X-ray sources with the GOODS-S catalog results in 200 matches, where we require a distance $<$1.5 arcsec for a successful match. Of these, 138 have spectroscopic redshifts (including all quality flags). This means that only $\sim$130 X-ray sources are assigned photometric redshifts (of a total of $\sim$32\,500 objects in the GOODS-S catalog) and we therefore expect the overall impact of these sources to be small. We further test the possible impact of objects with SEDs dominated by QSO/AGN contribution by including an additional set of template SEDs also covering AGNs and QSOs when deriving the photometric redshifts. For this purpose, we use template SEDs taken from the SWIRE SED library (Polletta et al. 2007). Our extended template set includes our six original templates and seven additional templates, the latter representing Seyfert galaxies, type 1 and type 2 QSOs as well as composite AGNs and starburst galaxies. 

The photometric redshifts derived with this extended set of template SEDs show very similar results as with the original set of SEDs. For the full spectroscopic sample including the 91 X-ray objects, we find $\sigma_{zc}$=0.039 for both the original template set and the extended template set. The fraction of outliers is slightly higher for the extended set, 3.2\% compared to 2.9\% for the original set. For the 91 X-ray sources themselves, we derive a scatter $\sigma_{zc}$=0.051 with 11.0\% outliers for the extended set and $\sigma_{zc}$=0.048 with 13.2\% outliers for the original set. 

From the above discussion, we conclude that there is no gain in accuracy when including templates covering AGNs and QSO SEDs when deriving the photometric redshifts using the GOODZ code. Contributing to this is the fact that normal galaxies dominate the GOODS-S field. The number of X-ray sources for which spectroscopic redshift are not available is also expected to be very small. Detailed measurements of photometric redshifts particularly aimed at the X-ray-selected sources is performed by Salvato et al. (2009) in the COSMOS field and by Luo et al. (2010) in the Chandra Deep Field South (covering GOODS-S).

\section{COMPARISON WITH OTHER RESULTS} 

When comparing published values of the photometric redshift errors (e.g., scatter between spectroscopic and photometric redshifts) between different surveys, one has to take into account that the results depend on the spectroscopic sample, spectral sampling and depths of filter set, as well as the way the errors are estimated and the treatment and definition of the outliers. Extremely accurate photometric redshift have been reported for the COSMOS survey (Ilbert et al. 2009). For a large sample of spectroscopic redshifts at $i^+_{\rm AB}<$22.5 and $z_{\rm spec}\lsim$1.5, they find $\sigma_{\rm NMAD'}=0.007$. Contributing to the success of these photometric redshifts is the large number of available filters (35 narrow band, intermediate band  and broadband) and spectroscopic redshifts used to ``train'' the template SEDs. While these filters, in particular the narrower bands, mainly cover optical wavelengths, we expect the accuracy of the COSMOS photometric redshifts to be best at $z\lsim$1.5 and at magnitudes where galaxies are detected in most of the available filters. At higher redshifts ($1.5<z<3$), Ilbert et al. find $\sigma_{\rm NMAD'}=0.054$~with $\sim$20\% outliers. Using the same definition of $\sigma_{\rm NMAD'}$ as Ilbert et al. (see below), we find over the same redshift interval a similar scatter $\sigma_{\rm NMAD'}=0.055$, but with fewer outliers, $\sim$8.3\% outliers (when including galaxies $i^+_{\rm AB}<25$). At faint magnitudes ($24<i^+_{\rm AB}<25$) and lower redshifts ($z\lsim$1.5), the COSMOS survey reports $\sigma_{\rm NMAD'}=0.053$~with $\sim$20\% outliers. For a comparable magnitude range and same redshift range, we find $\sigma_{\rm NMAD'}=0.043$~and $\sim$1.9\% outliers. Therefore, we conclude that the GOODZ photometric redshifts compare well with the COSMOS results in the redshift and magnitude ranges of interest in this investigation. For this comparison, we have used the definition $\sigma_{\rm NMAD'}$~from Ilbert et al. (2009), i.e., $\sigma_{\rm NMAD'}=1.48\times$median($|\Delta z|/(1+z_{\rm spec})$), which differs slightly from the definition adopted in Section 4. 

To make an even more direct comparison between results, one ideally wants to use the same galaxies and spectroscopic redshifts. We therefore turn to two surveys also probing GOODS-S for which we can use an identical spectroscopic sample

\subsection{FIREWORKS and GOODS-MUSIC}
Here, we make a direct comparison between the photometric redshifts derived in this investigation and the results in two publicly available photometric redshift catalogs: the GOODS-MUSIC catalog (Grazian et al. 2006; Santini et al. 2009) and the FIREWORKS catalog (Wuyts et al. 2008). 

The GOODS-MUSIC catalog contains objects selected in either the ACS $z$~band, the ISAAC $K_s$~band or the IRAC channel ch2 and includes photometric redshifts based on the VIMOS $U$~band (an earlier and less deep version of the data compared to the data used here), the four ACS bands, the three ISAAC bands, and the four IRAC channels. In addition, two other shallower $U$-band images from the ESO 2.2 m WFI camera, each with slightly different filter passbands, designated $U_{35}$~and $U_{38}$, are included. The photometry is derived using the PSF matching code ConvPhot (De Santis et al. 2007), which similar to the TFIT code used here, takes advantage of the high spatial resolution of the ACS data to measure accurate colors even in crowded regions. The GOODS-MUSIC version 2 catalog (Santini et al. 2009) used here lists 18,657 objects, for which photometric redshifts are given for 14,938 objects.  The photometric redshifts are based on a template fitting code described in Giallongo et al. (1998) and Fontana et al. (2000). To compare results, we match catalogs and extract a list of objects with spectroscopic data quality flag=1 that are common to both catalogs. This results in a sample of 1072 objects. We consider a successful match if the separation between the objects in the two catalogs is less than 0.5 arcsec. The resulting scatter, including galaxies with all magnitudes, is $\sigma_z$=0.14 for GOODZ and  $\sigma_z$=0.18 for GOODS-MUSIC. After excluding outliers, the scatter becomes  $\sigma_{zc}$=0.045 for GOODZ and $\sigma_{zc}$=0.055 for GOODS-MUSIC. The fractions of outliers for the two catalogs are 1.8\% and 3.1\%, respectively. Comparisons between the photometric and spectroscopic redshifts for these catalogs are shown in the left panels in Figure \ref{fig13}. 

The FIREWORKS catalog is an ISAAC $K_s$-selected catalog that includes photometry and photometric redshifts for 6307 objects over an area of 138 arcmin$^2$. Similar to our catalog, the FIREWORKS includes the ACS bands, the ISAAC bands, and the IRAC channels. In addition, the FIREWORKS catalog also includes ESO 2.2 m/WFI $U_{38}$-, $B$-, $V$-, $R$-, and $I$-band photometry. The photometric redshifts are calculated using the EAZY code (Brammer et al. 2008), which is based on the template fitting technique.

To compare results with FIREWORKS, we match catalogs using the above criteria, resulting in 1020 spectroscopic objects with quality flag=1, common between the two catalogs. We find a scatter of $\sigma_z$=0.092 for GOODZ and $\sigma_z$=0.12 for FIREWORKS, which decreases to  $\sigma_{zc}$=0.038 for GOODZ and $\sigma_{zc}$=0.036 for FIREWORKS after excluding outliers (2.3\% and 2.0\%, respectively). We show the comparison between the photometric and spectroscopic redshifts in the right panels of Figure \ref{fig13}.

In summary, the study in this section shows that different photometric redshift procedures, when applied on the same sample, give consistent results with high quality photometric redshifts. The catalog derived here using the GOODZ code does, however, include photometric redshifts for a significantly larger sample of objects compared to the GOODS-MUSIC and FIREWORKS catalogs. The GOODS-MUSIC catalog covers $\sim$93\% of the area we cover, while including in total $\sim$57\% as many objects (18,657 versus 32,508) and $\sim$46\% as many photometric redshifts (14,938 versus 32,505). Contributing to this difference is the use of ACS data from release v1.0 in GOODS-MUSIC while we in this investigation use the deeper version 2.0 of the data. The FIREWORKS catalog covers an area that is $\sim$90\% of the area covered here, although not fully overlapping with our area. The number of objects included is $\sim$20\% as many (6308 versus 32,508) as in our catalog. This large difference is mainly due to the ISAAC $K_s$-band selection used in the FIREWORKS catalog which preferentially selects red IR luminous objects and misses many blue objects detectable in the optical with ACS. The larger FWHM of ISAAC compared to ACS may also merge nearby objects that in an ACS-selected catalog are listed as individual sources. Other differences are that FIREWORKS uses ACS data version 1.0 (we use version v2.0) and ISAAC data version 1.5 (while we use the latest version 2.0). Comparison between GOODZ and GOODS-MUSIC $z$-band number counts are shown in the left panel of Figure \ref{fig14}. The right panel shows $K_s$-band number counts, including also FIREWORKS. The figure illustrates the fainter limits reached in this investigation.

We should note that besides the increase in the number of objects due to the larger area covered here (+7-10\%) compared to the GOODS-MUSIC and FIREWORKS catalogs, most of the additional objects are relatively faint with low S/N which may affect the quality of the photometric redshifts. Even though there is no significant increase in scatter to $m_z$=25.5 (Figure \ref{fig10}), this may not hold at fainter limits. Also, the fraction of outliers is expected to increase at fainter magnitudes.

\subsection{TFIT Versus SExtractor}
To compare our results from the TFIT magnitudes with those based on more ``traditional'' SExtractor (Bertin \& Arnouts 1996) magnitudes, we have created an alternative catalog as described in Section 2.2. We derive photometric redshifts for the SExtractor catalog using an aperture with 1.5 arcsec diameter for the $U$~band, the ACS bands, and the ISAAC bands. For IRAC, we use 3 arcsec aperture magnitudes due to the larger PSF for these data. We use the same iterative procedure as described above. This process will adjust the observed fluxes for necessary aperture corrections introduced by different aperture sizes, by matching fluxes with spectral templates using the sample of spectroscopic redshifts. 

We compare the photometric redshift estimates from the TFIT and SExtractor catalogs using galaxies with known spectroscopic redshifts in common between the two catalogs. The resulting scatter for the SExtractor catalog is $\sigma_{zc}$=0.046 and $\sigma_{zc}$=0.045 for the full sample ($N$=1271) and for $m_z<$24.5 ($N$=1123), respectively. For the same selection, using the TFIT catalog, we find $\sigma_z$=0.039 and $\sigma_z$=0.038 for the two selections, respectively. The outlier fractions are 7.2\% and 3.9\% for the SExtractor catalog compared to 3.6\% and 2.1\% for the TFIT catalog. Therefore, both the scatter and the fraction of outliers are higher for the SExtractor catalog compared to the TFIT catalog. 
One important difference between the two catalogs which could contribute to the results is that the TFIT method gives an estimate of the flux in all the bands that cover the area as given by the detection in the $z$~band, even though in some cases only an upper limit can be derived. For the SExtractor catalog, there is always a detection or an upper limit for the four ACS bands and the three ISAAC bands since the photometry for these bands is derived using SExtractor in ``dual image mode''. Only when a particular filter does not cover the full area of the detection band will there be a non-detection so that the particular filter cannot be used in the photometric redshift fitting. For the $U$~band and IRAC channels, the situation is different since for these bands we coordinate match objects to the ACS $z$-band-selected catalog, using a 1 arcsec matching radius. If there is no match between the $z$-band-selected objects and the $U$-band/IRAC catalogs, this could be due to the latter not covering the full area of the detection filter. However, it could also be that the flux is below the detection limit or blending may have caused absence of a match. Since we cannot distinguish between the latter two scenarios, we always exclude the non-matching filter in the photometric redshift calculation. To quantify this difference, we note that 75\% of the galaxies in the full spectroscopic sample are detected in equal numbers of filters in both catalogs, while 21\% are detected in more filters in the TFIT compared to the SExtractor catalog. The remaining 4\% are detected in more filters in the SExtractor catalog. Note that the spectroscopic sample has a relatively bright magnitude limit, which means that objects are, in most cases, detected in all available filters. However, at fainter magnitudes we expect objects to be undetected in an increasing number of filters. Here, the TFIT method gives valuable information by providing upper limits.

\section{COMPARISON BETWEEN THE PHOTOMETRIC REDSHIFT AND DROPOUT SELECTION TECHNIQUES} 
The dropout technique is a robust way to select high-redshift galaxies. Steidel et al. (1996) first used the technique to select Lyman break galaxies (LBGs) at $z\sim3$. Using different selection criteria, large samples of dropout galaxies in the redshift range $z\sim$3--6 have since been identified (Giavalisco et al. 2004b; Papovich et al. 2004; Bunker et al. 2004; Dickinson et al. 2004; Ouchi et al. 2004; Bouwens et al. 2006, 2007; Yoshida et al. 2006; Oesch et al. 2007) and searches for even higher redshift objects ($z\gsim$7) have been conducted (Yan \& Windhorst 2004; Bouwens et al. 2009; Capak et al. 2009; Ouchi et al. 2009; Castellano et al. 2010). 

Here, we compare Lyman break color selection of high-redshift galaxies to photometric redshifts. The photometric redshift distribution for LBGs can give insights about the redshift distribution and foreground contamination of LBG color selection. Comparing the LBG color selected samples to the overall $z$-band-selected population of galaxies with photometric redshifts in the same general range gives an idea of the efficiency and completeness of LBG selection. In particular, we focus on $B$-, $V$-, and $i$-band dropouts. To select the dropout samples, we use galaxy photometry from the SExtractor catalog (using the publicly available ACS-$z$-selected version 2.0 catalog\footnote{http://archive.stsci.edu/prepds/goods/}) in order to be consistent with previous works that base color selection criteria on SExtractor magnitudes. However, for the ACS bands in particular, the difference in colors based on TFIT and SExtractor magnitudes is marginal.

When selecting dropout galaxies with color criteria, the expected redshift distribution is characterized by a peak redshift and a standard deviation (Giavalisco et al. 2004b). In this investigation, we use the expected distribution of $B$-band and $V$-band dropout galaxies as described in detail in S. Salimbeni et al. (2010, in preparation). In short, artificial galaxies are distributed over redshift ranges $1.9<z<7.5$~according to an assumed distribution of luminosity. To construct the SED of the artificial galaxy, we use a stellar population model with a Salpeter initial mass function, a constant star formation rate and age of 0.14 Gyr, taken from Bruzual \& Charlot (2003). We use an evolving luminosity function to distribute the simulated LBGs in luminosity and redshift space based on the results of Reddy \& Steidel (2009; redshift bins $1.9<z<2.7$~and $2.7<z<3.4$) and Bouwens et al. (2007; redshift bins $3.4<z<4.5,~4.5<z<5.4,~5.4<z<6.5,$~and $6.5<z<7.5$). For the size of the simulated galaxies, we use a lognormal distribution that depends on galaxy magnitude. Finally, extinction is randomly added using a Gaussian distribution with a mean $E(B-V)=0.15$~and a width $\sigma_{E(B-V)}=0.15$.

Apparent colors are thereafter calculated after applying a Calzetti et al. (2000) extinction law as well as intergalactic absorption (Madau 1995). The simulated galaxies are then added to the ACS images and the images are processed with SExtractor. The $B$-dropout and $V$-dropout selections are applied on the detected galaxies. The simulations give us both the expected redshift distribution of the dropout-selected sample and the completeness function, $C(z)$, where the latter is defined as the number of color-selected galaxies compared to the total number of detected galaxies in the simulations above some detection limit as a function of redshift. Note that the dropout selection technique is not expected to select all galaxies within a predefined redshift range and will therefore always be incomplete to some extent. However, this can be corrected for if the completeness function, $C(z)$, is known from either simulations or observations (preferably a large spectroscopic sample). In order to compare galaxy samples selected using dropout criteria with photometric redshift selection, we have to define redshift ranges corresponding to each dropout selection criteria. In this paper we divide the redshift space into three contiguous redshift bins covering the ranges 2.8$<z<$4.4, 4.4$<z<$5.5, and 5.5$<z<$6.8 for  $B$-, $V$-, and $i$-band dropouts, respectively. These ranges include a major fraction of the galaxies selected by the different criteria, as shown below. Finally, the comparisons made here are between the photometric redshift method and the dropout technique, without implying that either of them gives the correct answer when it comes to e.g., completeness and contamination of the sample of selected high-redshift galaxies.

\subsection{$B$-band Dropouts}
To investigate consistency between the redshifts for high-$z$~candidates from the GOODZ photometric redshift method and the dropout technique, we use a sample of LBGs, selected using color criteria derived specifically for the GOODS-S/ACS survey (Giavalisco et al. 2004b; Papovich et al. 2004). For the $B$-band dropouts we use the selection criteria from Papovich et al. (2004)
\begin{equation}
(B-V > 1.1)\wedge(B-V>V-z+1.1)\wedge(V-z<1.6).
\end{equation}
To be consistent with earlier work, we only include objects with S/N$\ge$5 within an isophotal aperture in the $z$~band. If there is a non-detection in the $B$~band, we use the (S/N)$_B$=1 limit to derive a lower limit for the ($B-V$) color. We use SExtractor isophotal apertures defined in the $z$-band image for the purpose of measuring colors for selecting dropouts in a similar way as previous works. Furthermore, in our selection we exclude objects that are flagged as either stars or point sources. In total, there are 2129 objects that satisfy the $B$-band dropout criterion. 

In the left panel of Figure \ref{fig15}, we plot the photometric redshift distribution for the dropout sample. Light gray histogram shows the distribution using the Papovich et al. (2004) selection criterion in Equation (7), while darker color shows the subsample selected using the more restrictive criterion used in Giavalisco et al (2004b). Black line shows the predicted distribution based on simulations. This figure demonstrates a reasonably good agreement between the Lyman break dropout selection criteria and photometric redshifts. Both the photometric redshift distribution and the expected distribution from the dropout simulations peak at $z\sim$3.7. There is, however, a small low-redshift secondary peak in the photometric redshift distribution in Figure \ref{fig15} centered at $z\sim$0.5, as well as galaxies in the wings of the main peak that falls outside the defined redshift range 2.8$<z<$4.4. Note that the position of the secondary peak is consistent with the redshift where the 4000 \AA~Balmer breaks are aliased with the Lyman break. In total, $\sim$23\% (483 of 2129) of dropout-selected objects have photometric redshifts outside $2.8<z<4.4$, with about half of these being at low-redshift $z<2$ and half in the wings of the main peak of the redshift distribution. Using the redshift distribution from the simulations, we find that approximately 11\% of the galaxies are expected to have redshifts in the wings of the distribution, fairly consistent with the photometric redshifts. It is important to note that since the simulations do not extend below $z=1.9$, we are not able to quantify how many dropout-selected galaxies would fall in the secondary peak. Another simplification with the simulations that could affect comparisons is the use of a single galaxy SED to represent the high-redshift galaxies. However, at these redshifts, most detected galaxies should be consistent with being star-forming galaxies. This compares well with the result that $>$96\% of the photometric redshift selected galaxies in the redshift bin has a best-fitting galaxy type of a late-type star-forming galaxy. These issues with the present simulations and will be addressed in future work (S. Salimbeni et al. 2010, in preparation).

Without complete spectroscopic information, it is not possible to directly determine if the galaxies falling outside the redshift range are true contaminants of the dropout sample, consisting mostly of low-redshift galaxies, or if these are outliers with wrong photometric redshifts. Using the spectroscopic sample, we find that $\sim$20\% of the objects selected as dropout galaxies have a spectroscopic redshift outside the $2.8<z<4.4$ range (21 out of 103, including all quality flags), similar to what the photometric redshifts indicate. Vanzella et al. (2009) present spectroscopic follow-up of $B$-band dropouts (using the same color selection criteria as applied here) and find that 2 out of 48 objects have redshifts outside the predicted range, while Popesso et al. (2009) estimate a contamination fraction of $\sim$25\% for dropout-selected galaxies. The number of dropout objects with unexpectedly low photometric redshifts in our sample is therefore within the range of expected contamination fractions, estimated from the spectroscopic samples. Also, of the 483 dropout-selected objects with photometric redshifts outside the $2.8<z<4.4$ range, there exist spectroscopic redshifts for 11 objects. A majority of these, 9 of 11 (9 of 10 if excluding the lowest quality spectra), also have spectroscopic redshifts outside the redshift range. This suggests that the majority of the contamination is real and not driven by false photometric redshift determinations. 

Furthermore, the contamination factor should also depend on the limiting magnitude of the sample as well as the selection criteria used. For the latter, there is a trade-off between the number of selected dropout galaxies and the expected contamination. For example, the more restrictive selection criteria used in Giavalisco et al. (2004b), selects 1384 dropout galaxies, i.e., a subsample consisting of $\sim$35\% fewer objects compared to the selection used above. At the same time, however, the contamination fraction derived from the photometric redshifts drops to 15\%, compared to 23\% found for the less restrictive criteria. The galaxies selected with the latter criteria are plotted as the dark shaded histogram in Figure \ref{fig15}.

We also plot (Figure \ref{fig15} -- right panel) the $z$-band magnitude distribution for galaxies selected as $B$-band dropouts using both Papovich et al. (2004) and Giavalisco et al. (2004b) selection criteria. This distribution illustrates the faintness of the dropout sample, showing that the majority of the galaxies selected are significantly fainter than those in current spectroscopic samples, e.g., compare with Figure \ref{fig2}. 

To investigate the magnitude dependence of the contamination factor, we divide the galaxies selected with the Papovich criteria to four magnitude bins, each including $\sim$500 dropout galaxies. The median magnitudes for these bins are $m_z$=26.0, 27.0, 27.6, and 28.2, respectively (using SExtractor MAG$_{\rm ISO}$). From brighter to fainter bins we find, respectively, that 18.8\%, 21.6\%, 20.3\%, and 31.0\% of the dropout galaxies have photometric redshifts outside the $2.8<z<4.4$~range. This indicates that the contamination fraction is relatively independent of magnitude except at the very faintest limits where there is an indication of an increase of contaminants. However, if we look only at the fraction of galaxies in the low-redshift peak ($z<1$), we find fractions 9.5\%, 10.8\%, 7.6\%, and 4.0\%. This indicates that the overall increase of galaxies outside the range in the faintest bin is not due to a misidentification between the Lyman and the Balmer breaks, but instead due to a general increase in photometric redshift errors, shifting more galaxies outside the defined redshift range.

Besides the contamination fraction, it is important to understand the completeness of the dropout selection, i.e., what fraction of the total number of galaxies within a defined redshift range is expected to be selected with the dropout criteria. Based on photometric redshifts, we find a total of 3318 galaxies in the range $2.8<z<4.4$, of which 1646 have colors according to the dropout criteria, suggesting a completeness of $\sim$50\%. Using the simulated completeness function, we find that 33\% of the simulated galaxies within the redshift range are selected with the $B$-band dropout criteria, indicating that the simulations somewhat underestimates the completeness compared to the photometric redshifts. These results are illustrated in Figure \ref{fig16}, which shows the completeness as a function of redshift for both the photometric redshift selected sample and the simulations. There is a good agreement between the location and shape of the distributions, however, as noted above the completeness is higher for the photometric redshift sample.

There is of course a direct relation between the contamination fraction and the completeness. Choosing a narrower photometric redshift range for the $B$-band dropouts will increase the contamination and at the same time also increase the completeness. E.g., adopting the redshift range 3.44$<z<$4.12, used by Giavalisco et al. (2004b), we find a contamination fraction of 53\% (compared to 23\% when using the wider redshift range). At the same time, the completeness for the narrower redshift range is 76\% compared to 50\% for the wider range.

To illustrate the dropout selection, we show in Figure \ref{fig17} a color--color diagram where the $B$-band dropout galaxies reside on the top left part of the plot. Dropout galaxies with photometric redshifts in the range $2.8<z<4.4$~are shown as crosses. Triangles outside the region defined by the selection criteria show galaxies within the redshift range that are not selected as dropouts. We note that these galaxies have a similar ($V-z$) color as the dropout-selected candidates and have a median ($B-V$) color that is $\sim$0.4 mag bluer than the color selection limits. We also find that of the total number of galaxies in the redshift range $2.8<z<4.4$, not selected by the $B$-band dropout criteria, less than 2\% (25 of 1672) are selected as $V$-band dropouts in the range $4.4<z<5.5$~(selection discussed below). Finally, $\sim 23\%$ of the dropout-selected galaxies that were found to have photometric redshifts outside the above range are shown as filled circles inside the dropout selection criteria region in Figure \ref{fig17}. As expected, most of these are close to the selection boundary. 

Previously, Popesso et al. (2009) have estimated that the $B$-band selection criteria from Giavalisco et al. (2004b) results in a completeness of $\sim$80\%, in good agreement with the 76\% found here for the same selection. In contrast, using $Spitzer$~galaxies selected in IRAC channel 2 at 4.5 $\mu$m, Mancini et al. (2009) estimate that dropout methods would miss $\sim$80\% of the $z\ge$3.5 galaxies. The reason for this high incompleteness is that the majority of the $Spitzer$~selected $z\ge$3.5 galaxies are too faint to be detected in the optical passbands and therefore, are absent from the dropout catalogs.

It should be noted that a relative good agreement between the photometric redshifts and dropout selection is expected since both methods use colors to locate the Lyman break and from this derive the redshift. The main difference is that the dropout technique uses the optical bands only, while the photometric redshift method also uses available NIR and MIR data to constrain the overall shape of the galaxies SEDs. To quantify this, we find that of the $B$-band dropout-selected galaxies, 56\% are detected in both ISAAC and IRAC, while an additional 34\% are detected in either ISAAC or IRAC, leaving only 10\% of the dropout galaxies without IR detection.

\subsection{$V$-band Dropouts}
The $V$-band dropout selection criterion is based on a shorter wavelength baseline, only using an upper limit for the $B$~band. This could increase the risk of contamination and incompleteness. Also, these dropouts are at higher redshifts and therefore, on average, fainter with higher photometric uncertainties, which may reduce the accuracy of the photometric redshift estimates. Vanzella et al. (2009) use spectroscopic follow-up of dropout-selected galaxies and found $\sim$ 4\% contamination for $B$-band dropouts (2 of 48), which increases to 11\% (4 of 36) for $V$-band dropout. To investigate the relation between the photometric redshifts and dropout selections, we use the $V$-band dropout selection criteria from Giavalisco et al. (2004b),  
\begin{equation}
[(V-i)>1.5+0.9\times(i-z)]\vee[(V-i)>2.0]\wedge
(V-i)\ge1.2 \wedge (i-z)\le1.3 \wedge [{\rm (S/N)}_B<2]
\end{equation}
In addition, we again require (S/N)$_z\ge$5 to include an object in the sample. Figure \ref{fig18} shows the photometric redshift distributions for the $V$-band dropouts. As expected, the contamination fraction is higher compared to the $B$-band selection. For the $V$~band, we find that 29\% (140 of 490) of the selected objects have photometric redshifts outside the expected range $4.4<z<5.5$. More than half of these are found in a secondary peak centered at $z\sim$0.9, while a smaller fraction (4\% of the total) is found in the wings of the main peak. Again, the location of the secondary peak is consistent with the position where the 4000 \AA~Balmer breaks are aliased with the Lyman break. These results are also consistent with the $\sim$25\%~contamination fraction by low-redshift galaxies in $V$-band dropout samples found by Popesso et al. (2009). Using the spectroscopic sample, we find that 8 out of 34 ($\sim$24\%) of the objects selected as dropout galaxies have spectroscopic redshifts outside the defined redshift range, which is in agreement with the numbers found above. Spectroscopic redshifts only exists for 7 dropout-selected objects that have photometric redshifts outside the range $4.4<z<5.5$. Of these, 3 also have spectroscopic redshifts outside the range while 4 are inside the range. However, 2 of these 4 have photometric redshifts barely outside the $z=5.5$ limit and are therefore not true outliers (i.e., both have $|\Delta z|/(1+z_{\rm spec})<0.03$). These results indicate that a majority of the low-redshift contamination is real, but the uncertainty in this number is larger compared to the $B$-band selection.

Figure \ref{fig18} also shows the expected distribution based on simulations. The agreement is good with both distributions covering the same redshift range. However, the photometric redshifts have a slightly narrower and more peaked distribution compared to the simulations. The fraction of galaxies from the simulations that are in the wings of the main peak is 14\%.

Calculating the completeness, we find that 56\% (350 of 628) of the galaxies with photometric redshifts in the range $4.4<z<5.5$~are selected by the $V$-band dropout criteria. In addition, 20 galaxies in this range that are not selected with the $V$-band criteria are selected by either the $B$-band or $i$-band criteria. Using the simulations, we find a completeness of 49\%, in good agreement with the photometric redshifts. Figure \ref{fig16} shows that the redshift dependences of the completeness also agree well between methods.

\subsection{$i$-band Dropouts}
The $i$-band dropout selection criterion is based on a single $i-z$~color together with upper limits for the $B$~and $V$~band (Dickinson et al. 2004):
\begin{equation}
(i-z)\ge 1.3 \wedge {\rm (S/N)}_B<2 \wedge {\rm (S/N)}_V<2.
\end{equation}
With only a single color, we expect an even higher contamination fraction compared to the $V$-band dropouts. Consequently, Vanzella et al. (2009) found 18\% contamination for spectroscopically confirmed $i$-band dropouts (6 of 34), higher than for both $B$- and $V$-band dropouts. Figure \ref{fig19} shows the photometric redshift distributions for the $i$-band dropouts (requiring (S/N)$_z\ge$5). We find that 49\% (103 of 212) of the $i$-band selected galaxies have photometric redshifts outside the expected range of $5.5<z<6.8$. A major fraction of these make up the secondary low-redshift peak centered at $z\sim$1.2, consistent with the position expected if the two main breaks are aliased. Calculating the completeness we find that $\sim$87\% (109 of 126) of galaxies with photometric redshift in the given redshift range are selected by the dropout selection criteria. Of the non-selected objects with photometric redshift $5.5<z<6.8$, 10 galaxies are selected by the $V$-band dropout selection technique and none by the $B$-band selection. 

Of the dropout-selected galaxies, there exist 22 spectroscopic redshifts of which 2 have a redshift outside the range $5.5<z<6.8$. This contamination fraction is smaller than suggested by the photometric redshifts, but small statistics and selection effects in the spectroscopic sample could affect the difference. Spectroscopic redshifts exist for two of the dropout-selected galaxies residing in the low photometric redshift peak at $z\sim$1.2. One of these has a high spectroscopic redshift consistent with being a dropout galaxy, while the other has a spectroscopic redshift outside the range. These statistics are too small to draw any conclusions from.     

A higher expected contamination fraction for the $i$-band dropout sample is a consequence of the single color selection criterion. In an early analysis of shallower GOODS data using 3 epochs of $HST$~observations, compared to $\sim 10$~epochs for the final GOODS version 2 data products, Dickinson et al. (2004) found a robust sample of 5 $i$-band dropouts with (S/N)$_z \geq 10$. Applying the same color selection and S/N limit to the deeper ACS version 2 data, we find 35 objects, of which 22 have photometric redshifts in the range $5.5<z<6.8$, indicating a $\sim$37\% contamination. Of the $i$-dropout-selected galaxies using the higher S/N cut, we have spectroscopic redshifts for 14 objects. All these objects have redshift within the $5.5<z<6.8$~range, suggesting that this selection is relatively safe. None of the objects in the low-redshift peak has an available spectroscopic redshift.

For fainter objects (5$\le$(S/N)$_z\le$10) Dickinson et al. use simulations to estimate the contamination fraction and find $\sim$45\% contamination in the $i$-band dropout sample. This is similar to the $\sim$51\% contamination we find using the photometric redshifts for objects selected in the same (S/N)$_z$~range.

\section{THE GOODS-S PHOTOMETRIC REDSHIFT CATALOG}
Of a total of 32,508 objects in the $HST$/ACS $z$-band-selected photometric catalog, we calculate photometric redshifts for 32,505, excluding only a few objects that do not have photometry in at least three bands, which we require to calculate photometric redshifts. In Figure \ref{fig20}, we plot the photometric redshift distribution using three different magnitude limits $m_z<$24 (black), $m_z<$25 (dark gray), $m_z<$26 (light gray). Besides the photometric redshifts (both the integrated $z_{\rm phot}$~(Equation (3)) and peak value $z_{\rm peak}$), our catalog also lists the 68\% and 95\% confidence intervals for the photometric redshifts, as well as the best-fitting spectral type. The number of objects flagged as stars based on colors in the catalog is 845, however, we still give the photometric redshift for these objects since non-stellar objects may be flagged as stars, in particular at faint magnitudes where photometric errors may be large. We also flag 464 objects as point sources based on a surface brightness--magnitude relation. The catalog is matched to 2875 spectroscopic redshifts (see Section 2.3). Furthermore, of the $\sim$270 X-ray objects in Alexander et al. that fall within the GOODS-S ACS-$z$ selected area, we find 200 matches when requiring a separation $<$1.5 arcsec. Finally, of the 64 radio sources in the catalog from Afonso et al. (2006), we find 53 matches using the same matching criterion. Matched X-ray and radio sources are flagged in the photometric redshift catalog.

The photometric redshift catalog will be made public in a simultaneous release of catalogs for both GOODS-S and GOODS-N (T. Dahlen et al. 2010, in preparation). 
\section{CONCLUSIONS AND SUMMARY}
We have calculated photometric redshifts for the GOODS South field using VIMOS $U$~band, ACS $B$, $V$, $i$, $z$~bands, ISAAC $J$, $H$, $K_s$~bands, and four IRAC channels ch1 to ch4 centered on 3.6$\mu$m, 4.5$\mu$m, 5.8$\mu$m, and 8.0$\mu$m. Photometry is derived using TFIT which provides consistent photometry in the different filters regardless of the PSF size differences between instruments. We have used a ``training'' set of spectroscopic redshifts to compare the observed fluxes and the fluxes predicted from our set of template SEDs. Using the differences between observed and predicted fluxes, we have calculated and applied mean offsets for each filter as well as rest wavelength dependent corrections of the template SEDs.

Our final catalog covers an area of 153 arcmin$^2$ and contains photometric redshifts for 32,505 objects. Our main conclusions are:
\begin{itemize}   
\item{Comparing our photometric redshifts to the available spectroscopic data, we measure an overall scatter is $\sigma_{zc}\sim$0.040 with an outlier fraction of 3.7\% for the full sample, while for $m_z<$24.5 we find $\sigma_{zc}\sim$0.039 with 2.1\% outliers. The systematic bias of the photometric redshifts is only $\Delta z/(1+z)=-0.006$~for the full sample and -0.005 for the brighter subsample with $m_z<$24.5.}
\item{The scatter between spectroscopic redshifts and photometric redshifts is comparable to the best results previously published for GOODS-S (FIREWORKS and GOODS-MUSIC). However, the current catalog is deeper and provides photometric redshifts for a significantly larger sample of galaxies.}
\item{Based on color--color criteria, we flag objects with colors consistent with being stars. Objects with point source morphology are flagged using surface brightness - magnitude relation.}
\item{Including the $U$~band is important for accurate photometric redshifts, especially at low-redshift $z<0.3$ and at $z\sim$2.}
\item{Including infrared photometry is crucial when deriving photometric redshifts, especially at redshifts $z>1.2$.}
\item{The presence dust emission (e.g., PAH features) at rest wavelengths $\gsim 3\mu$m affects the fluxes in the IRAC filters. The template SEDs have to account for this to decrease the scatter in the photometric redshifts.}
\item{Deriving redshifts using TFIT based photometry instead of SExtractor photometry improves the photometric redshifts.}
\item{Studying the photometric redshift for objects selected as $B$-band dropouts, we find that 77\% of the sample have redshift in the expected range 2.8$<z<$4.4. Of the total number of galaxies with photometric redshift in this range, we find that the $B$-band dropout selection criterion selects 50\% of the objects. This is reasonably consistent with what is expected from dropout selection. For all dropout selections we use a limit (S/N)$_z\ge$5. }
\item{For $V$-band dropouts, which are selected using a shorter wavelength baseline compared to $B$-band dropouts, we find as expected somewhat larger differences between photometric redshifts and dropout selections. Of the $V$-band dropouts we find that 71\% of the sample has a redshift in the range 4.4$<z<$5.5. Of the galaxies in this redshift range, the $V$-band selection finds 56\% of the objects.} 
\item{The $i$-band dropouts are selected on basically only one color, which should lead to high uncertainties.
Consequently, only 51\% of the $i$-band dropouts have a photometric redshift in the range 5.5$<z<$6.8. The fraction of the galaxies in this high-redshift range that is selected by the $i$-band dropout criterion is 87\%.} 

\end{itemize}
\acknowledgments{
Based on observations made with the NASA/ESA $Hubble Space Telescope$, obtained at the Space Telescope Science Institute, which is operated by the Association of Universities for Research in Astronomy, Inc., under NASA contract NAS 5-26555. These observations are associated with programs GO-9352, GO-9425, GO-9583, GO-9728, GO-10189, GO-10339, and GO-10340. 
Observations have been carried out using the Very Large Telescope at the ESO Paranal Observatory under Program ID(s): LP168.A-0485.
This work is based in part on observations made with the $Spitzer Space Telescope$, which is operated by the Jet Propulsion Laboratory, California Institute of Technology, under a contract with NASA. Support for this work was provided by NASA through an award issued by JPL/Caltech.
M.N. acknowledges the financial contribution from contract ASI I/016/07/0 and from the PRIN INAF ‚ ``{\it A deep VLT and LBT view of the Early Universe: the physics of high-redshift galaxies}''. We thank the anonymous referee for valuable comments and suggestions.
}

\clearpage
\appendix
\section{SYSTEMATIC ERROR ESTIMATES}
\begin{figure}
\epsscale{0.9}
\plotone{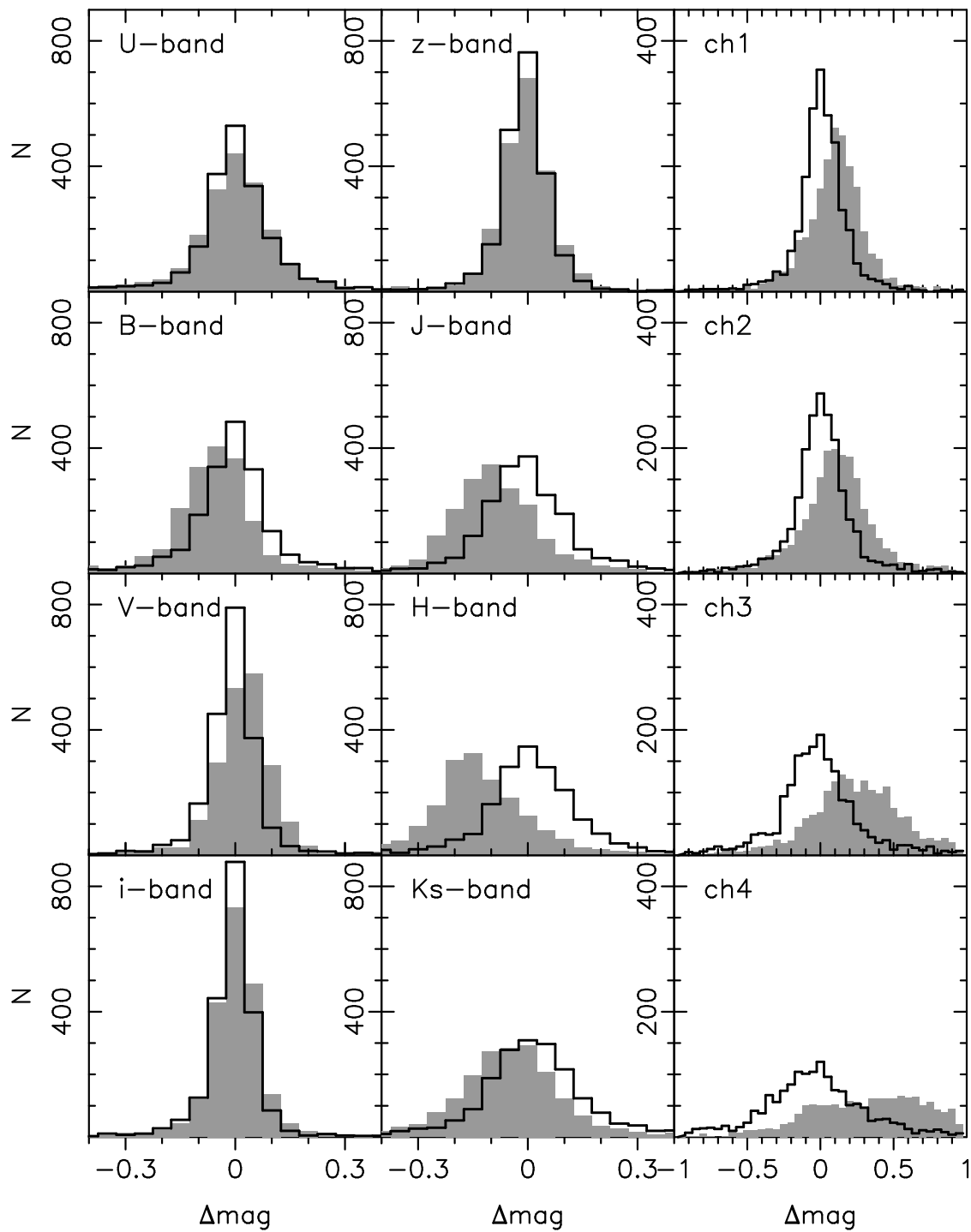}
\caption{Difference between the predicted and observed (from template fitting) magnitudes for the sample of 2288 galaxies with spectroscopic redshifts (data quality 1 and 2). Gray histograms show the distribution before any corrections were applied. Black lines show distributions after correcting magnitude for the offset shifts found and using updated template SEDs, as discussed in Sections 3.2.1, 3.2.3, and 3.2.5.
\label{fig6}}
\end{figure}

\begin{figure}
\epsscale{0.7}
\plotone{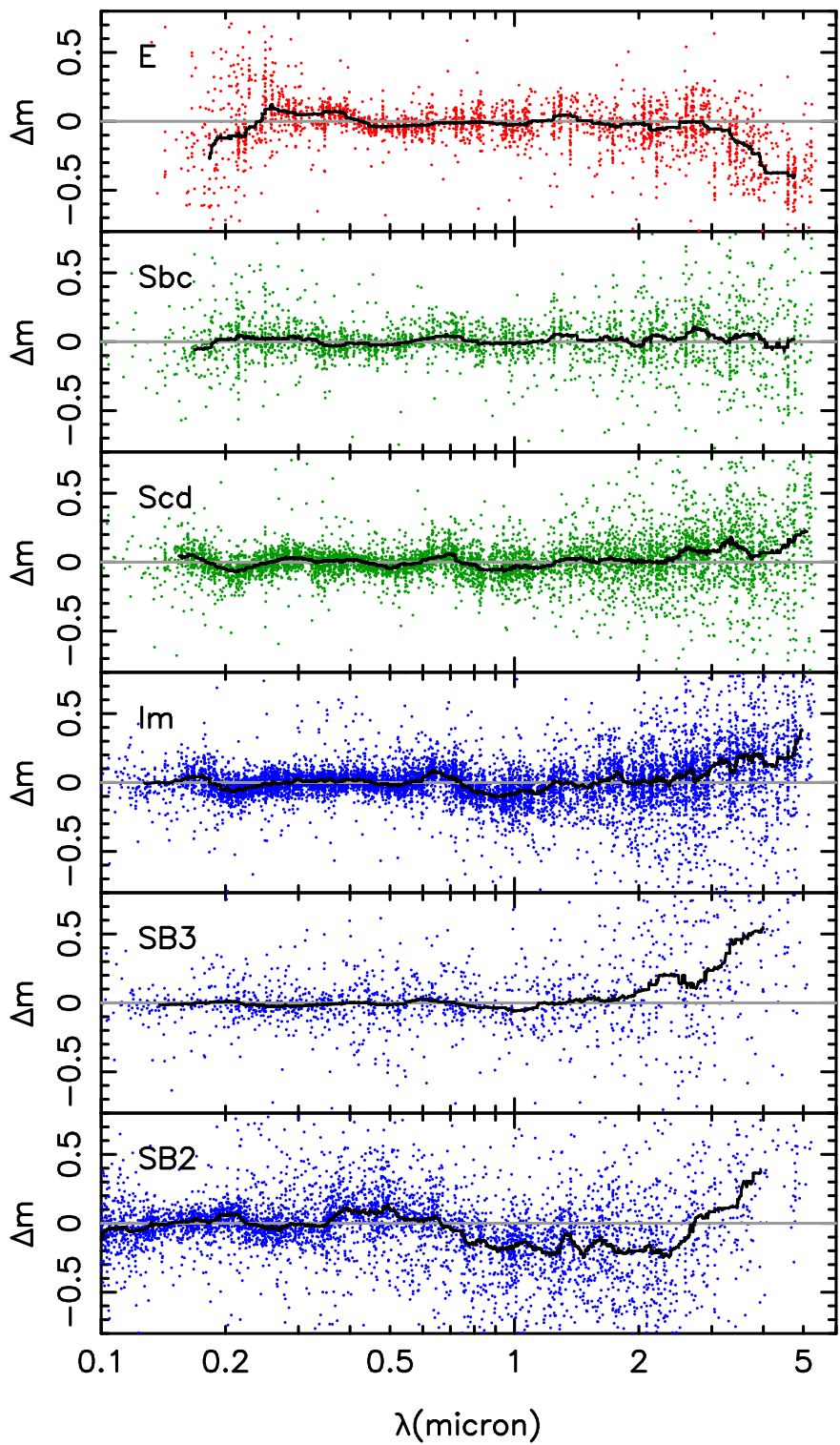}
\caption{Offsets between observed photometry and photometry derived from template SEDs as a function of rest-frame wavelength for the six different galaxy types.  A positive offset indicates that the observed flux is brighter compared to that expected from the template SED. Solid curves show the median of the offset.
\label{fig7}}
\end{figure}

\begin{figure}
\epsscale{0.7}
\plotone{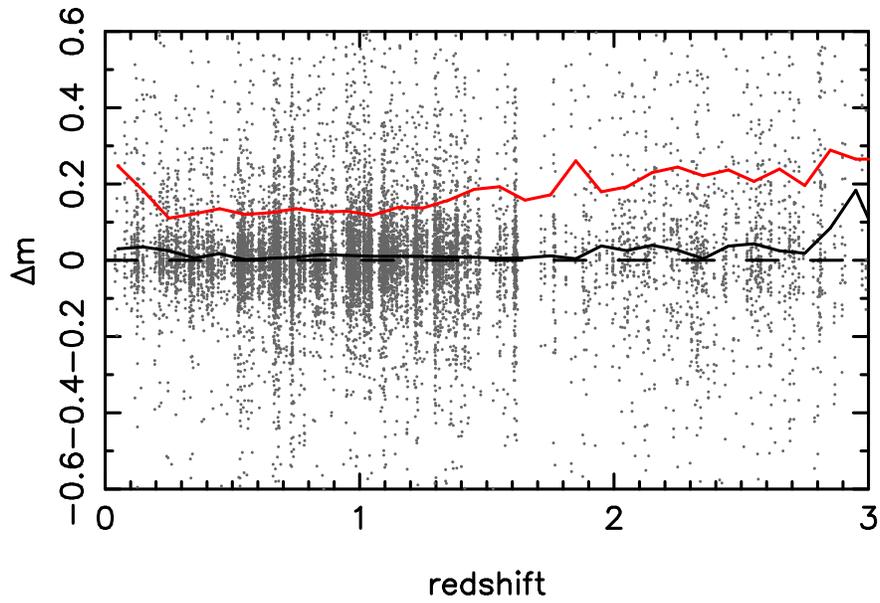}
\caption{Offsets between observed photometry and photometry derived from template SEDs as a function of redshift. Lower curve shows the median of the offset, while the upper curve shows the rms.
\label{fig8}}
\end{figure}

\begin{figure}
\epsscale{0.7}
\plotone{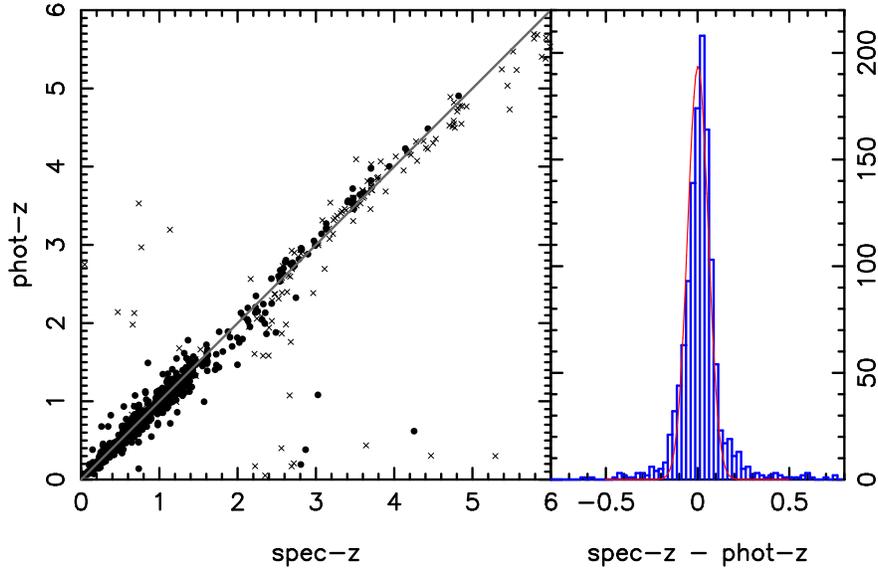}
\caption{Left panel: photometric vs. spectroscopic redshift. Black dots show objects with $m_z<$24.5 (1118 objects), while crosses show fainter objects (162). Right panel shows a histogram over the difference between spectroscopic redshifts and photometric redshifts ($z_{\rm spec}-z_{\rm phot}$) for the full sample of 1280 objects. Overplotted is a best-fit Gaussian distribution with $\sigma$=0.056. For the redshift normalized distribution ([$z_{\rm spec}-z_{\rm phot}$]/[1+$z_{\rm spec}$]), we get $\sigma$=0.038.
\label{fig9}}
\end{figure}

\begin{figure}
\epsscale{0.7}
\plotone{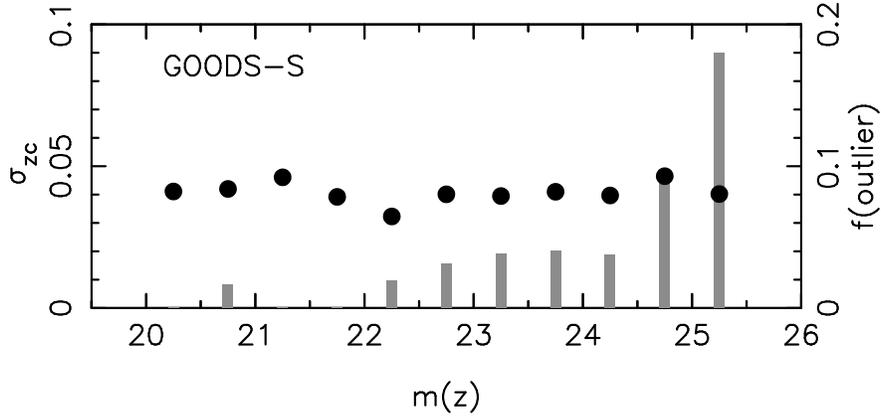}
\caption{Photometric redshift scatter ($\sigma_{zc}$) as a function of magnitude is shown with black dots and scaling on left-hand $y$-axis. The outliers are excluded when estimating $\sigma_{zc}$ values. Histograms show the fraction of outliers as a function of magnitude (scaling on right-hand $y$-axis).
\label{fig10}}
\end{figure}

\begin{figure}
\epsscale{0.7}
\plotone{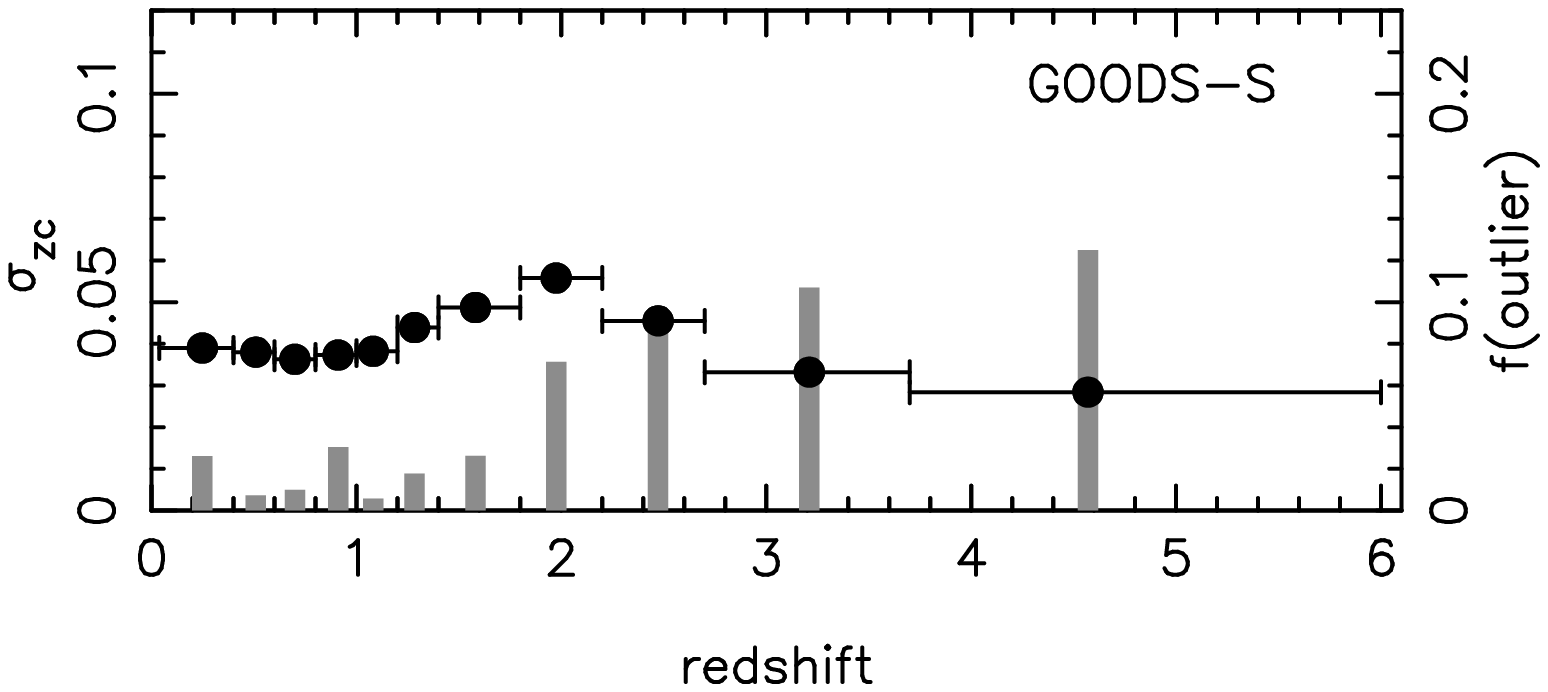}
\caption{Photometric redshift scatter ($\sigma_{zc}$) as a function of redshift is shown with black dots and scaling on left-hand $y$-axis. Histograms show the fraction of outliers as a function of redshift (scaling on right-hand $y$-axis). The magnitude limit applied is $m_z<24.5$. 
\label{fig11}}
\end{figure}

\begin{figure}
\epsscale{0.7}
\plotone{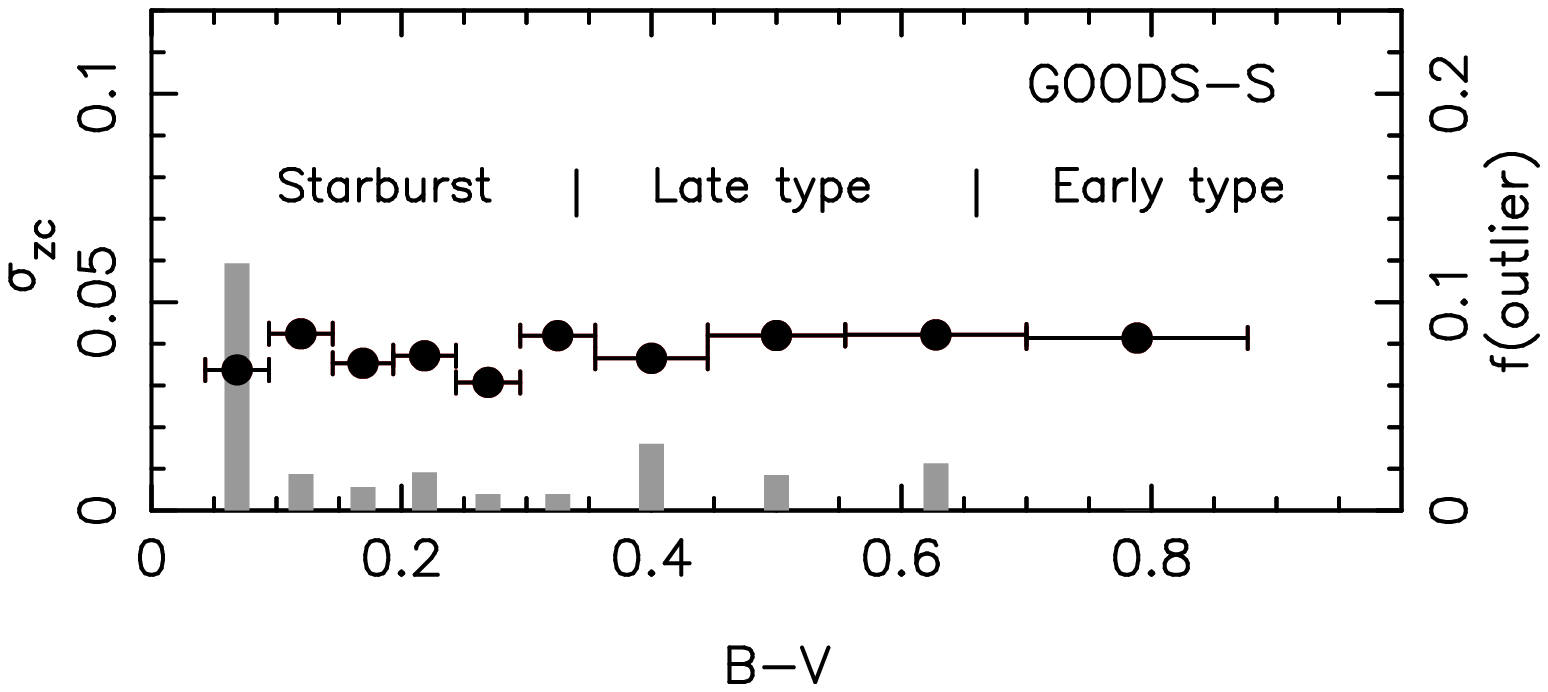}
\caption{Photometric redshift scatter ($\sigma_{zc}$) as a function of galaxy color is shown with black dots and scaling on left-hand $y$-axis. Histograms show the fraction of outliers as a function of redshift (scaling on right-hand $y$-axis). The magnitude limit applied is $m_z<24.5$. 
\label{fig12}}
\end{figure}

\begin{figure}
\epsscale{0.9}
\plotone{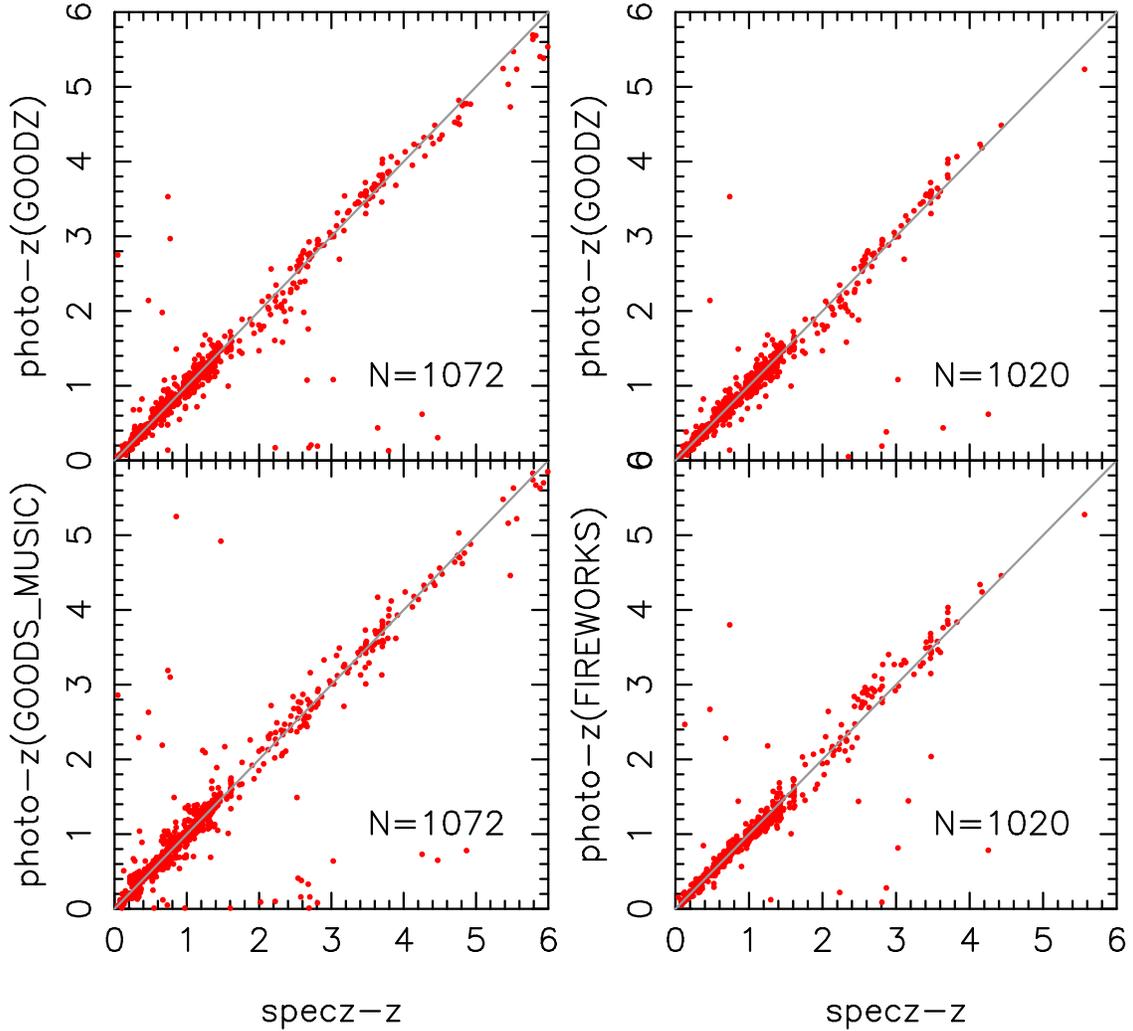}
\caption{Left panels: comparison between photometric and spectroscopic redshifts from the GOODZ catalog (top) and the GOODS-MUSIC catalog (bottom) for 1072 objects with spectroscopic redshift (quality flag=1). Right panels: comparison between photometric and spectroscopic redshifts from the GOODSZ catalog (top) and the FIREWORKS catalog (bottom) for 1020 objects with spectroscopic redshift.
\label{fig13}}
\end{figure}

\begin{figure}
\epsscale{0.7}
\plotone{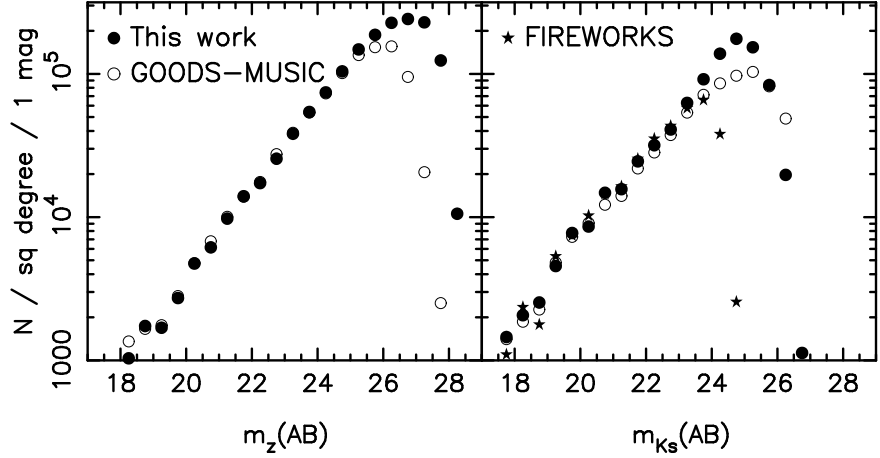}
\caption{Left panel: GOODS-S number counts in the $z$~band for this investigation (filled circles) and for GOODS-MUSIC (open circles). Right panel: number counts in the $K_s$~band including also the FIREWORKS catalog.  
\label{fig14}}
\end{figure}

\begin{figure}
\epsscale{0.7}
\plotone{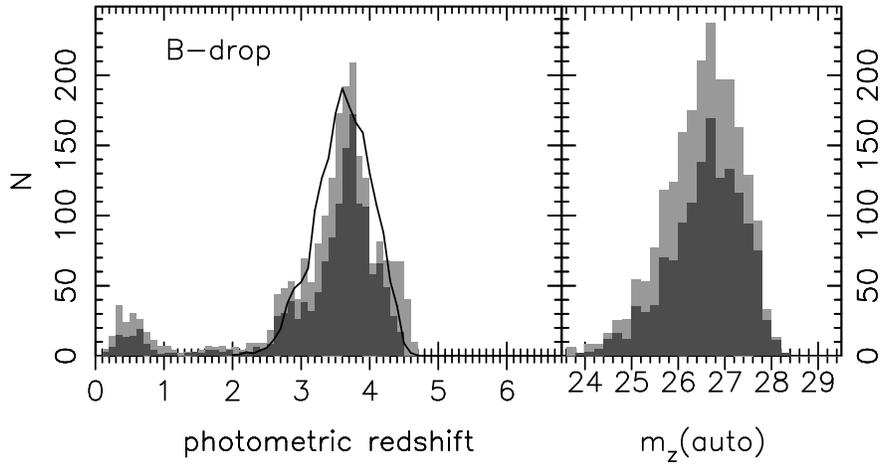}
\caption{Left panel: photometric redshift distribution for galaxies selected as $B$-band dropouts in GOODS-S using the color selection in Papovich et al. (2004; light gray). Also shown, is the subsample selected with the more restrictive selection criteria in Giavalisco et al. (2004b; dark gray). Solid black line shows the expected distribution based on simulations (S. Salimbeni et al. 2010, in preparation). Right panel: the $z$-band magnitudes of the dropout sample.
\label{fig15}}
\end{figure}

\begin{figure}
\epsscale{0.7}
\plotone{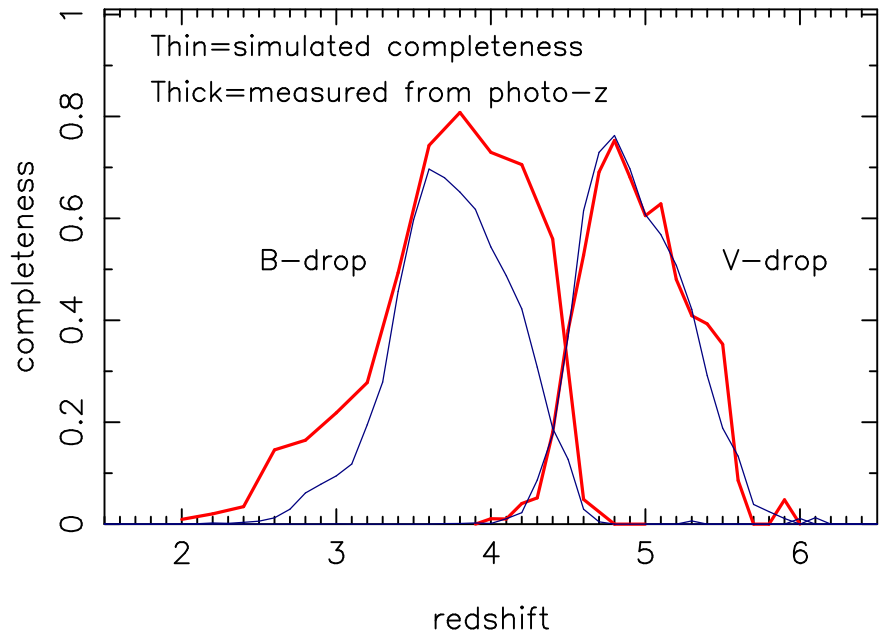}
\caption{Completeness of the $B$-dropout and $V$-dropout samples as derived from the photometric redshifts (thick lines) and simulations (thin lines). The completeness is defined as the fraction of galaxies selected using color criteria compared to the total number of galaxies at each redshift. For both samples we use a selection criterion (S/N)$_z\ge$5).
\label{fig16}}
\end{figure}

\begin{figure}
\epsscale{0.7}
\plotone{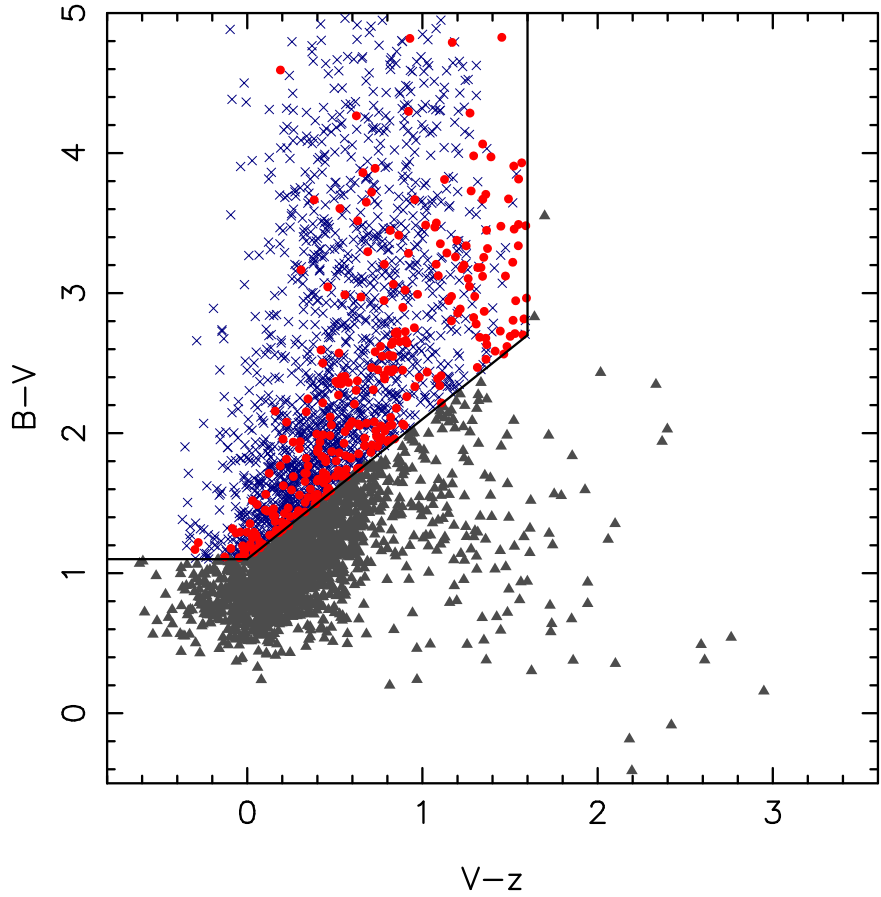}
\caption{Color--color diagram for the $B$-band dropouts. Galaxies selected by the dropout criteria resides in the top left part of the plot. Crosses mark $B$-dropout-selected objects with photometric redshifts in the range $2.7<z<4.1$, while filled circles are objects with photometric redshifts outside the redshift range. Triangles are objects with photometric redshifts inside this redshift range that are not selected by the dropout criteria. 
\label{fig17}}
\end{figure}

\begin{figure}
\epsscale{0.7}
\plotone{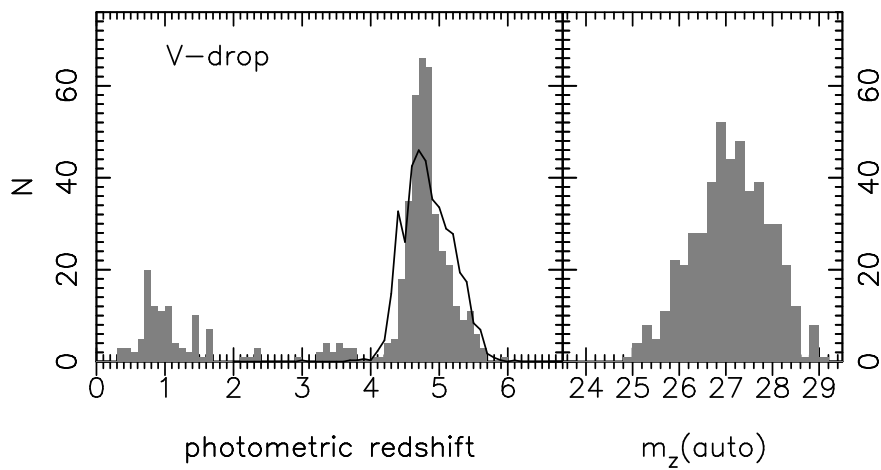}
\caption{Photometric redshift distribution for galaxies selected as $V$-band dropouts in GOODS-S using the color selection in Giavalisco et al. (2004b). Solid black line shows the expected distribution based on simulations (S. Salimbeni et al. 2010, in preparation). Right panel shows the $z$-band magnitudes of the dropout sample.
\label{fig18}}
\end{figure}

\begin{figure}
\epsscale{0.7}
\plotone{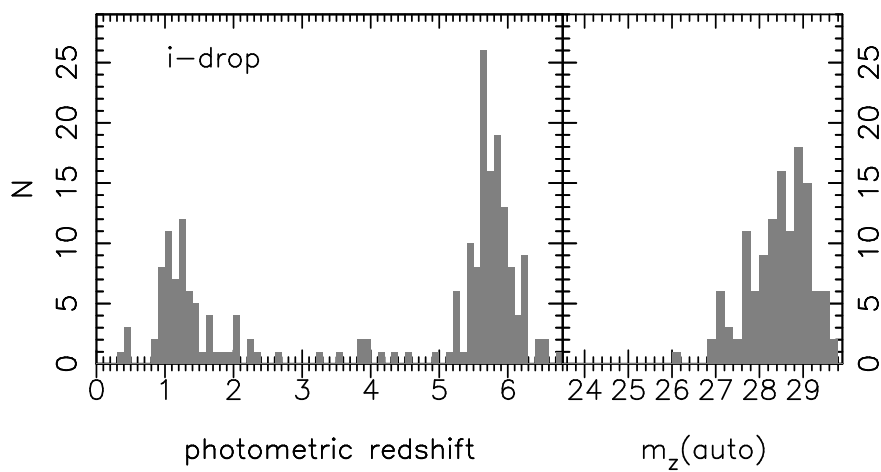}
\caption{Photometric redshift distribution for galaxies selected as $i$-band dropouts in GOODS-S using the color selection in Giavalisco et al. (2004b). Right panel shows the $z$-band magnitudes of the dropout sample.
\label{fig19}}
\end{figure}

\begin{figure}
\epsscale{0.7}
\plotone{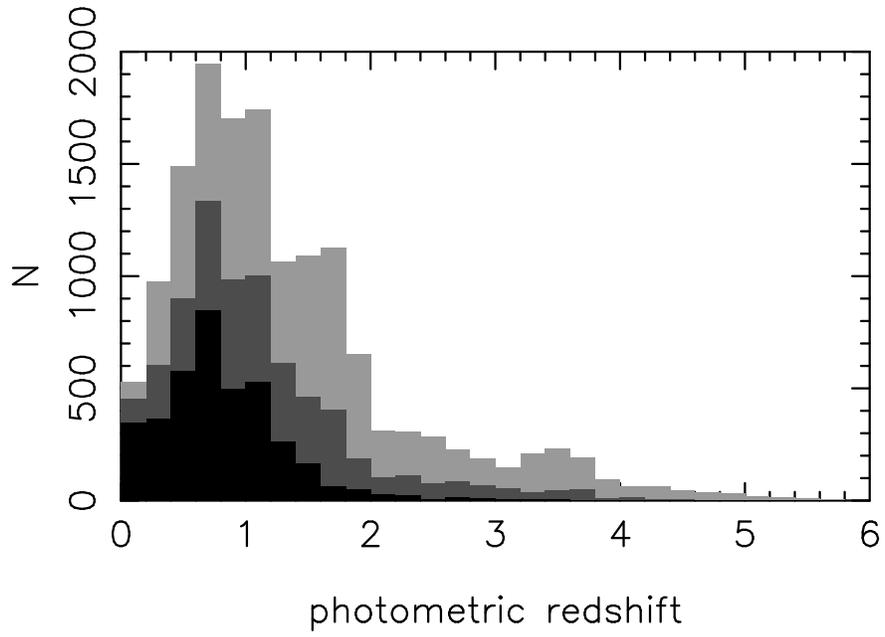}
\caption{Photometric redshift distribution for the GOODS-S catalog for three different magnitude limits: $m_z<$24 (black), $m_z<$25 (dark gray), $m_z<$26 (light gray). 
\label{fig20}}
\end{figure}

The scatter between the observed photometry and the template photometry shown with the black lines in Figure \ref{fig6} should in an ideal case represent the statistical errors in the measured fluxes. In a real case, however, several factors may add systematic errors that increase the scatter. This includes uncorrected photometric zero-point errors, calibration errors (e.g., flat-field and dark corrections), insufficient knowledge of filter response functions and the detector quantum efficiency. Furthermore, even if the template SEDs well matches the shapes of the real galaxies, there will always be some deviation between the discrete template set and the true ``continuous'' set of real galaxies. These deviations will increase if the template set does not well represent the true galaxy shapes. In this investigation we do, however, decrease that risk by training our template SEDs using the spectroscopic sample. In the chi-square fitting performed by the photometric redshift code, the errors included in Equation (1) should take into account the total uncertainty in the fit between the observed magnitude and the template SED. Including only statistical photometric errors underestimates the errors and may cause systematic effects in the photometric redshift fitting. 

\begin{figure}
\epsscale{0.7}
\plotone{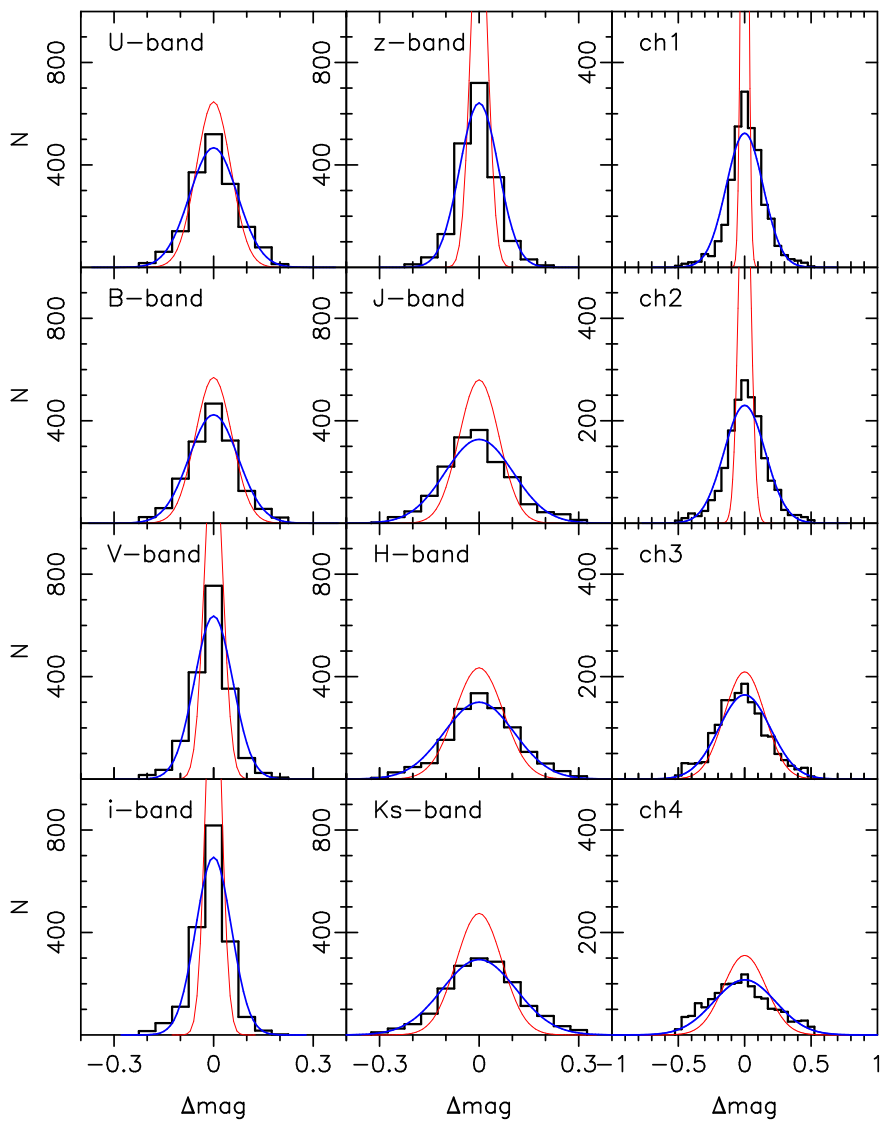}
\caption{Histograms show residuals between observed photometry and template photometry for a sample of objects with spectroscopic redshifts. Overplotted with thick (blue) lines are Gaussians with widths given by the rms scatter of the residuals. Thin (red) lines show the distribution of photometric errors. The scatter in the residuals is larger than the photometric errors in all bands. 
\label{fig21}}
\end{figure}

We investigate the size of the total systematic errors not accounted for in the photometric errors given by our photometry catalog to get an estimate of the additional smoothing errors that should be added to the photometric errors in the chi-square fitting procedure. Figure \ref{fig21} shows histograms over the residuals between observed and template SED magnitudes for the spectroscopic sample using a magnitude cut $m_z<$24.5. The over-plotted thick lines show Gaussian distributions with a width given by the rms of the distribution of residuals. Thin lines show Gaussian functions with a width that equals the rms of the statistical photometric errors in our catalog. Note the different scaling for IRAC bands in the rightmost column. The figure shows that for the ACS $V$, $i$, and $z$~bands and for IRAC channels ch1 and ch2, the photometric errors are significantly smaller than the distribution of residuals. However, all bands do show larger residuals than photometric scatter. This is expected since systematic effects are not included in the photometric scatter. We find that adding (in quadrature) ``smoothing'' errors $\sigma_{\rm add}$=0.05 to $U$~band and ACS bands, $\sigma_{\rm add}$= 0.1 to the ISAAC bands, and $\sigma_{\rm add}$=0.2 to the IRAC channels, results in a decreased scatter between the spectroscopic and photometric redshifts as well as a significant decrease in the fraction of outliers. Adding these errors to the pure photometric error distributions plotted with the thin lines in Figure \ref{fig21}, makes these distributions consistent with the distributions of the observed scatter (thin red lines). The larger total errors ensures us that the systematic errors are not underestimated when performing the $\chi^2$~fitting.

%% Generally speaking, only the figure captions, and not the figures
%% themselves, are included in electronic manuscript submissions.
%% Use \figcaption to format your figure captions. They should begin on a
%% new page.

\clearpage

%% No more than seven \figcaption commands are allowed per page,
%% so if you have more than seven captions, insert a \clearpage

%% There must be a \figcaption command for each legend. Key the text of the
%% legend and the optional \label in curly braces. If you wish, you may
%% include the name of the corresponding figure file in square brackets.
%% The label is for identification purposes only. It will not insert the
%% figures themselves into the document.
%% If you want to include your art in the paper, use \plotone.
%% Refer to the on-line documentation for details.

\clearpage

\clearpage

\clearpage
\clearpage

\begin{deluxetable}{lr}
\tabletypesize{\scriptsize}
\tablewidth{0pt}
\tablecaption{Median Offsets Between Template and Measured Fluxes for the Filters Used in this Investigation Before any Corrections are Applied.}
\tablecolumns{2}
\tablehead{
\colhead{}Filter & Initial offset  
}
\startdata
$U$~(VIMOS)  & --0.003  \\
$B$~(ACS)    & --0.051  \\
$V$~(ACS)    &  0.028  \\
$i$~(ACS)    &  0.004  \\
$z$~(ACS)    &  0.001  \\ 
$J$~(ISAAC)  & --0.086  \\
$H$~(ISAAC)  & --0.122  \\
$K_s$~(ISAAC) & --0.021  \\
ch1 (IRAC) &  0.132  \\
ch2 (IRAC) &  0.130  \\
ch3 (IRAC) &  0.284  \\
ch4 (IRAC) &  0.486
\enddata
\tablecomments{Offsets are given in magnitudes even though the fitting is done in flux space.
}
\label{table1}
\end{deluxetable}

\begin{deluxetable}{llccccccr}
\tabletypesize{\scriptsize}
\tablewidth{0pt}
\tablecaption{Photometric redshift results}
\tablecolumns{9}
\tablehead{
\colhead ~Selection~~~~~~~ & m$_{\rm lim}$ & Redshift & N$_{\rm spec}$  & $\sigma_z$ & $\sigma_{zc}$& $\sigma_{\rm NMAD}$ & f(OL) & bias$_z$
}
\startdata
All filters   & All         & All & 1280 & 0.135 & 0.040 & 0.035  & 3.7\%  & --0.006 \\
\hline
All filters   & $m_z<$24.5  & All & 1118 & 0.062 & 0.039 & 0.034   & 2.1\% & --0.005 \\
Excluding $U$ & $m_z<$24.5  & All & 1118 & 0.074 & 0.044 & 0.040   & 3.4\% & --0.003 \\
Excluding IR  & $m_z<$24.5  & All & 1118 & 0.175 & 0.047 & 0.044   & 5.5\% &  0.004 \\
Excluding NIR & $m_z<$24.5  & All & 1118 & 0.076 & 0.041 & 0.035   & 2.5\% & --0.005\\ 
Excluding MIR & $m_z<$24.5  & All & 1118 & 0.083 & 0.041 & 0.037   & 2.6\% & --0.006\\
\hline
All filters   & $m_z<$24.5& $z<0.3$ & 94 & 0.058 & 0.043 & 0.040   & 2.1\% &  0.006\\
Excluding $U$ & $m_z<$24.5& $z<0.3$ & 94 & 0.134 & 0.064 & 0.068   & 13.8\%&  0.052\\
\hline
All filters   & $m_z<$24.5& $2.0<z<2.3$ & 12 & 0.049 & 0.049 & 0.049   & 0.0\% &  --0.026\\
Excluding $U$ & $m_z<$24.5& $2.0<z<2.3$ & 12 & 0.122 & 0.075 & 0.055   & 16.7\%&  --0.081\\
\hline
All filters   & $m_z<$24.5& $z>1.2$ & 223 & 0.099 & 0.044 & 0.038   & 4.5\% & --0.015\\
Excluding IR  & $m_z<$24.5& $z>1.2$ & 223 & 0.116 & 0.051 & 0.048   & 9.0\%& 0.012\\
\hline
With X-ray sources  & $m_z<$24.5& All & 1209 & 0.075 & 0.039 & 0.035   & 2.9\% & --0.004\\
Only X-ray sources & $m_z<$24.5& All & 91 & 0.167 & 0.048 & 0.048   & 13.2\%& 0.013
\enddata
\tablecomments{Definitions are given in Section 4.
}
\label{table2}
\end{deluxetable}

\end{document}